\documentclass[cleveref,english, thm-restate]{lipics-v2021}
\usepackage[T1]{fontenc}
\usepackage[utf8]{inputenc}
\usepackage{amsmath,amsfonts,amssymb,amsthm}
\usepackage[dvipsnames,svgnames, table]{xcolor}
\usepackage{graphicx}
\usepackage{enumerate}
\usepackage[disable]{todonotes}
\usepackage{float}
\hypersetup{hidelinks,breaklinks}

\usepackage{diagbox}
\usepackage{array} %to center content in tabular environments
\usepackage{arydshln} %dashed lines
\usepackage{makecell}

\newif\ifshort
\shortfalse
\newcommand{\shortlong}[2]{\ifshort #1\else #2\fi}
\newcommand{\shortapp}[1]{\shortlong{the long version}{Appendix~\ref{#1}}}

\usepackage{mathcommand}
\usepackage[notion, quotation, electronic]{knowledge}

\usepackage{tikz}
\usetikzlibrary{automata, snakes, positioning, arrows,arrows.meta,calc,decorations,decorations.markings,math,shapes.geometric}

\tikzset{
	>=stealth',%Black traingles
	-={stealth',ultra thick,scale=3} % makes the arrow heads bold
	node distance=1cm, % specifies the minimum distance between two nodes.
	every state/.style={thick}, % sets the properties for each node
	initial text=$ $, % sets the text that appears on the start arrow
}

\title{On the Minimisation of Deterministic and History-Deterministic Generalised (co)B\"uchi Automata}
\titlerunning{On the Minimisation of (History-)Deterministic Generalised (co)B\"uchi Automata}

\author{Antonio Casares}{LaBRI, Université de Bordeaux, France \and University of Warsaw, Poland \and \url{https://antonio-casares.github.io/}}{antoniocasares@mimuw.edu.pl}{https://orcid.org/0000-0002-6539-2020}{}
\author{Olivier Idir}{IRIF, Université Paris-Cité, France}{}{}{}
\author{Denis Kuperberg}{LIP, CNRS, ENS Lyon, France \and \url{https://www.perso.ens-lyon.fr/denis.kuperberg/}}{denis.kuperberg@ens-lyon.fr}{https://orcid.org/0000-0001-5406-717X}{}
\author{Corto Mascle}{LaBRI, Université de Bordeaux, France \and \url{https://corto-mascle.github.io/}}{corto.mascle@labri.fr}{https://orcid.org/0009-0007-7976-7480}{}
\author{Keya Prakash}{University of Warwick, United Kingdom \and \url{https://keya-prakash.github.io}}{keya.prakash@proton.me}{https://orcid.org/0000-0002-2404-0707}{}

\authorrunning{A. Casares, O. Idir, D. Kuperberg, C. Mascle and K. Prakash} 
\Copyright{Antonio Casares, Olivier Idir, Denis Kuperberg, Corto Mascle and Keya Prakash}

\ccsdesc[500]{Theory of computation~Logic~Verification by model checking} %TODO mandatory: Please choose ACM 2012 classifications from https://dl.acm.org/ccs/ccs_flat.cfm 
\keywords{Automata minimisation, omega-regular languages, good-for-games automata}
\category{} %optional, e.g. invited paper
\nolinenumbers %uncomment to disable line numbering

\EventEditors{J\"{o}rg Endrullis and Sylvain Schmitz}
\EventNoEds{2}
\EventLongTitle{33rd EACSL Annual Conference on Computer Science Logic (CSL 2025)}
\EventShortTitle{CSL 2025}
\EventAcronym{CSL}
\EventYear{2025}
\EventDate{February 10--14, 2025}
\EventLocation{Amsterdam, Netherlands}
\EventLogo{}
\SeriesVolume{326}
\ArticleNo{22}

\newcommand\ap[1]{\todo[inline,size=\scriptsize,backgroundcolor=Goldenrod]{#1 - \textbf{Keya}}}

\let\ab\allowbreak
\mathchardef\hyphen=45 %Decimal

\newcommand{\set}[1]{\{#1\}}
\newrobustcmd{\pow}[1]{2^{#1}}
\newrobustcmd{\powplus}[1]{\kl[\powplus]{2^{#1}_{+}}}
\knowledge{\powplus}{notion}
\newrobustcmd{\partialF}{\mathrel{\kl[\partialF]{\rightharpoonup}}}
\knowledge{\partialF}{notion}

\newcommand{\size}[1]{|#1|}

\newcommand{\disjUnion}{\mathrel{\uplus}}
\newrobustcmd{\infEL}{\mathtt{Inf}}
\newrobustcmd{\finEL}{\mathtt{Fin}}

\newrobustcmd{\emptyword}{\kl[\emptyword]{\varepsilon}}
\knowledge{\emptyword}{notion}

\newrobustcmd{\prefix}{\mathrel{\kl[\prefix]{\sqsubseteq}}}
\knowledge{\prefix}{notion}
\newrobustcmd{\nprefix}{\mathrel{\kl[\nprefix]\sqsubset}}
\knowledge{\nprefix}{notion}

\newrobustcmd{\first}{\kl[\first]{\mathsf{first}}}
\knowledge\first{notion}
\newrobustcmd{\last}{\kl[\last]{\mathsf{last}}}
\knowledge\last{notion}

\newrobustcmd{\minf}{\kl[\minf]{\mathsf{Inf}}}
\knowledge{\minf}{notion}
\newrobustcmd{\mocc}{\kl[\mocc]{\mathsf{Occ}}}
\knowledge{\mocc}{notion}

\newcommand{\re}[1]{\xrightarrow{#1}}
\usetikzlibrary{calc,decorations.pathmorphing,shapes} %%

\newcounter{sarrow}
\newcommand\lrp[1]{%
	\stepcounter{sarrow}%
	\mathrel{\begin{tikzpicture}[baseline= {( $ (current bounding box.south) + (0,-0.5ex) $ )}]
			\node[inner sep=.5ex] (\thesarrow) {$\scriptstyle #1$};
			\path[draw,<-,decorate,
			decoration={zigzag,amplitude=0.7pt,segment length=1.2mm,pre=lineto,pre length=4pt}] 
			(\thesarrow.south east) -- (\thesarrow.south west);
	\end{tikzpicture}}%
}
\newcommand\lrpE{\lrp{\phantom{w.}}}

\newrobustcmd\Lang[1]{\kl[\Lang]{\mathcal{L}(#1)}}
\knowledge\Lang{notion}

\newrobustcmd{\init}{\mathrm{init}}
\newrobustcmd\autComposition{\mathrel{\kl[\autComposition]{\circ}}}
\knowledge\autComposition{notion}

\newrobustcmd\outRun{\kl[\outRun]{\mathsf{Output}}}
\knowledge\outRun{notion}
\newrobustcmd\colInf{\kl[\colInf]{\mathsf{colInf}}}
\knowledge\colInf{notion}

\newcommand{\colAut}{\mathsf{col}}

\newrobustcmd\resolvPath[1]{\kl[\resolvPath]{#1^*}}
\knowledge\resolvPath{notion}

\newrobustcmd\Unfold[1]{\kl[Unfold]{\mathsf{Unfold}(#1)}}
\knowledge\Unfold{notion}

\newrobustcmd{\initialAut}[2]{\kl[\initialAut]{#1^{#2}}}
\knowledge\initialAut{notion}

\newrobustcmd\safeA{\kl[\safeA]{\A_\mathsf{safe}}}
\knowledge\safeA{notion}

\newrobustcmd{\mout}{\kl[\mout]{\mathsf{Out}}}
\knowledge\mout{notion}
\newrobustcmd{\mIn}{\kl[\mIn]{\mathsf{In}}}
\knowledge\mIn{notion}

\newrobustcmd{\Buchi}{\kl[\Buchi]{\mathsf{B\"uchi}}}
\knowledge\Buchi[\BuchiCol]{notion}
\newrobustcmd{\coBuchi}{\kl[\coBuchi]{\mathsf{coB\"uchi}}}
\knowledge\coBuchi[\coBuchiCol]{notion}
\newrobustcmd{\genBuchi}{\kl[\genBuchi]{\mathsf{genB}}}
\knowledge\genBuchi[\genBuchiCol]{notion}
\newrobustcmd{\genCoBuchi}{\kl[\genCoBuchi]{\mathsf{genCoB}}}
\knowledge\genCoBuchi[\genCoBuchiCol]{notion}

\newrobustcmd{\genBuchiCol}[1]{\kl[\genBuchiCol]{\mathsf{genB}_{#1}}}
\newrobustcmd{\genCoBuchiCol}[1]{\kl[\genCoBuchiCol]{\mathsf{genCoB}_{#1}}}

\newrobustcmd{\autGenCoB}[1]{\kl[\autGenCoB]{\D^{\mathrm{coB}}_{#1}}}
\knowledge\autGenCoB{notion}

\newrobustcmd{\autGenB}[1]{\kl[\autGenB]{\D^{\mathrm{B}}_{#1}}}
\knowledge\autGenB{notion}

\newrobustcmd{\Res}[1]{\kl[\Res]{\mathsf{Res}(#1)}}
\knowledge{\Res}{notion}

\newrobustcmd{\quot}[2]{\kl[\quot]{#1^{-1}#2}}
\knowledge{\quot}{notion}

\newrobustcmd{\resStates}[1]{\kl[\resStates]{Q^{#1}}}
\newrobustcmd{\resStatesA}[2]{\kl[\resStatesA]{Q^{#1}_{#2}}}
\knowledge{\resStates}[\resStatesA]{notion}

\newrobustcmd{\resClass}[1]{\kl[\resClass]{[#1]}}
\knowledge{\resClass}{notion}
\newrobustcmd{\resClassLang}[1]{\kl[\resClassLang]{[#1]_L}}
\knowledge{\resClassLang}{notion}
\newrobustcmd{\resClassAut}[2]{\kl[\resClassAut]{[#1]_{#2}}}
\knowledge{\resClassAut}{notion}
\newrobustcmd{\restRes}[2]{\kl[\restRes]{\restr{#1}{#2}}}
\knowledge{\restRes}{notion}

\newrobustcmd\eqRes{\mathrel{\kl[\eqRes]\sim}}
\newrobustcmd\eqResAut[1]{\mathrel{\kl[\eqResAut]\sim_{#1}}}
\knowledge{\eqRes}[\eqResAut]{notion}
\newrobustcmd\eqSafe[1]{\mathrel{\kl[\eqSafe]\approx{#1}}}
\knowledge{\eqSafe}{notion}

\newrobustcmd\locAlph[1]{\kl[\locAlph]{\restr{\SS}{#1}}}
\knowledge{\locAlph}{notion}
\newrobustcmd\locLang[1]{\kl[\locLang]{\restr{L}{#1}}}
\knowledge{\locLang}{notion}
\newrobustcmd\localAut[1]{\kl[\localAut]{\restr{\A}{#1}}}
\newrobustcmd\localAutA[2]{\kl[\localAutA]{\restr{#1}{#2}}}
\knowledge{\localAut}[\localAutA]{notion}

\newrobustcmd\autMinAK[1]{\kl[\autMinAK]{\mathcal{A}_{#1}^{\mathsf{coB}}}}
\newrobustcmd\autMinAKL{\kl[\autMinAKL]{\mathcal{A}_{L}^{\mathsf{coB}}}}
\knowledge\autMinAK[\autMinAKL]{notion}
\newrobustcmd\autMinGenCoB{\kl[\autMinGenCoB]{\mathcal{A}_{L}^{\mathsf{genCoB}}}}
\knowledge\autMinGenCoB{notion}
\newrobustcmd\autMinGenCoBPI{\kl[\autMinGenCoBPI]{\mathcal{A}_{L}^{\mathsf{genCoB}}}}
\knowledge\autMinGenCoBPI{notion}

\newrobustcmd\safeComp[1]{\kl[\safeComp]{\mathsf{Safe}(#1)}}
\knowledge\safeComp{notion}
\newrobustcmd\safeLang[1]{\kl[\safeLang]{\L_{\mathsf{Safe}}(\A^{#1})}}
\newrobustcmd\safeLangA[2]{\kl[\safeLangA]{\L_{\mathsf{Safe}}(#1^{#2})}}
\knowledge\safeLang[\safeLangA]{notion}
\newrobustcmd\colouredsafeLangA[3]{\kl[\colouredsafeLangA]{\L_{\mathsf{#3-Safe}}(#1^#2)}}
\knowledge\safeLang[\safeLangA]{notion}

\newrobustcmd{\chromNum}{\kl[\chromNum]{\chi}}
\knowledge\chromNum{notion}

\newrobustcmd{\neighbourhood}[1]{\kl[\neighbourhood]{N[#1]}}
\knowledge\neighbourhood{notion}
\newrobustcmd{\neighbourhoodOpen}[1]{\kl[\neighbourhoodOpen]{n(#1)}}
\knowledge\neighbourhoodOpen{notion}

\newrobustcmd{\langVertex}[1]{\kl[\langVertex]{L_{#1}}}
\knowledge\langVertex{notion}
\newrobustcmd{\langGraph}{\kl[\langGraph]{L_G}}
\knowledge\langGraph{notion}

\newrobustcmd{\langClique}{\kl[\langClique]{L_{\exists\mathrm{noRep}}}}
\knowledge\langClique{notion}

\newrobustcmd{\FTwoLetters}{\kl[\FTwoLetters]{\F_G}}
\knowledge\FTwoLetters{notion}

\NewDocumentCommand{\pbMinGenBuchi}{}{\kl[\pbMinGenBuchi]{\normalfont{\textsc{\small{Minimisation of generalised B\"uchi automata}}}}}
\NewDocumentCommand{\pbMinGenCoBuchi}{}{\kl[\pbMinGenCoBuchi]{\normalfont{\textsc{\small{Minimisation of generalised coB\"uchi automata}}}}}
\knowledge\pbMinGenBuchi[\pbMinGenCoBuchi]{notion}

\NewDocumentCommand{\pbChromNum}{}{\kl[\pbChromNum]{\normalfont{\textsc{\small{Chromatic number}}}}}
\knowledge\pbChromNum{notion}
\NewDocumentCommand{\pbThreeCol}{}{\kl[\pbThreeCol]{\normalfont{\textsc{\small{3-colorability}}}}}
\knowledge\pbThreeCol{notion}

\newrobustcmd\Gtwo[1]{\kl[\Gtwo]{G_2(#1)}}
\knowledge\Gtwo{notion}

\newrobustcmd{\Stab}[1]{\kl[\Stab]{\textsf{Stab}}(#1)}
\knowledge\Stab{notion}

\newrobustcmd{\adj}[1]{\kl[\adj]{\textsf{adj}}(#1)}
\knowledge\adj{notion}

\newrobustcmd{\Lstab}[1]{\kl[\Lstab]{L_{#1}^{\textsf{stab}}}}
\knowledge\Lstab{notion}
\newcommand\restr[2]{
		#1\hspace{-1.1mm}\upharpoonright_{\hspace{-0.2mm}#2} 
}

\newcommand\restrOLD[2]{{% we make the whole thing an ordinary symbol
		\left.\kern-\nulldelimiterspace % automatically resize the bar with \right
		#1 % the function
		\littletaller % pretend it's a little taller at normal size
		\right|_{#2} % this is the delimiter
}}

\newcommand{\littletaller}{\mathchoice{\vphantom{\big|}}{}{}{}}

\newrobustcmd\inv[1]{#1^{-1}}

\newcommand{\NP}{\ensuremath{\mathsf{NP}}}
\newcommand{\NPc}{\ensuremath{\mathsf{NP}\hyphen\mathrm{complete}}}

\newcommand{\PTimeFull}{\ensuremath{\mathsf{PTIME}}}

\newrobustcmd{\FO}{\ensuremath{\mathrm{FO}}}
\newrobustcmd{\LTL}{\kl[\LTL]{\ensuremath{\mathrm{LTL}}}}

\newrobustcmd{\MSO}{\ensuremath{\mathrm{MSO}}}
\newrobustcmd{\SOneS}{\ensuremath{\mathrm{S1S}}}
\newrobustcmd{\STwoS}{\ensuremath{\mathrm{S2S}}}
\DeclareMathAlphabet{\mathpzc}{OT1}{pzc}{m}{it}

\newrobustcmd{\NN}{\mathbb{N}}
\newrobustcmd{\ZZ}{\mathbb{Z}}
\newrobustcmd{\QQ}{\mathbb{Q}}
\newrobustcmd{\RR}{\mathbb{R}}
\newrobustcmd{\CC}{\mathbb{C}}
\newrobustcmd{\WW}{\mathbb{W}}

\newrobustcmd{\I}{\mathcal{I}}
\newrobustcmd{\F}{\mathcal{F}}
\newrobustcmd{\D}{\mathcal{D}}
\newrobustcmd{\N}{\mathcal{N}}
\newrobustcmd{\G}{\mathcal{G}}

\renewcommand{\L}{\mathcal{L}}
\newrobustcmd{\M}{\mathcal{M}}
\newrobustcmd{\Q}{\mathcal{Q}}
\newrobustcmd{\C}{\mathcal{C}}
\newrobustcmd{\A}{\mathcal{A}}
\newrobustcmd{\B}{\mathcal{B}}
\newrobustcmd{\Z}{\mathcal{Z}}
\newrobustcmd{\R}{\mathcal{R}}
\newrobustcmd{\T}{\mathcal{T}}
\newrobustcmd{\U}{\mathcal{U}}
\newrobustcmd{\W}{\mathcal{W}}

\renewcommand{\O}{\mathcal{O}}

\renewcommand{\S}{\mathcal{S}}

\newrobustcmd{\kk}{\kappa}
\newrobustcmd{\uu}{\upsilon}
\newrobustcmd{\dd}{\delta}

\renewcommand{\ss}{\sigma}

\newrobustcmd{\rr}{\rho}

\renewcommand{\aa}{\alpha}

\newrobustcmd{\bb}{\beta}

\newrobustcmd{\oo}{\omega}
\newrobustcmd{\pp}{\varphi}

\renewcommand{\gg}{\gamma}
\newrobustcmd{\ee}{\varepsilon}

\renewcommand{\SS}{\Sigma}
\newrobustcmd{\GG}{\Gamma}
\newrobustcmd{\DD}{\Delta}

\knowledgerenewmathcommand\nu{\cmdkl{\LaTeXnu}}
\knowledgenewmathcommand\nuAcd{\cmdkl{\LaTeXnu}}
\knowledgerenewmathcommand\eta{\cmdkl{\LaTeXeta}}
\IfKnowledgePaperModeTF{
	}{
		\knowledgestyle{intro notion}{color={Dark Ruby Red}, emphasize}
		\knowledgestyle{notion}{color={Dark Blue Sapphire}}
		\hypersetup{
				colorlinks=true,
				breaklinks=true,
				linkcolor={}, % Links to sections, pages, etc.
				citecolor={}, % Links to bibliography
				filecolor={Blue Marine}, % Links to local file
				urlcolor={Blue Marine},
			}
	}

\knowledge{notion}
| notion
| definition

\knowledge{notion}
| $\omega $-word
| $\omega $-words
| infinite words
| infinite word

\knowledge{notion}
| word
| words

\knowledge{notion}
| empty word

\knowledge{notion}
| prefix

\knowledge{notion}
| factor

\knowledge{notion}
| partial mapping

\knowledge{notion}
| state complexity

\knowledge{notion}
| directed graph
| graph@directed
| graphs@directed

\knowledge{notion}
| undirected graph
| undirected 
| undirected graphs
| graph
| graphs
| finite undirected graph

\knowledge{notion}
| neighbour
| neighbours

\knowledge{notion}
| neighbourhood

\knowledge{notion}
| strict neighbourhood

\knowledge{notion}
| size@graph

\knowledge{notion}  
|  connected@und
|  connected

\knowledge{notion}
| simple@graph

\knowledge{notion}
| path
| paths

\knowledge{notion}
| strongly connected

\knowledge{notion}
| subgraph
| subgraphs

\knowledge{notion}
| strongly connected component
| SCC
| strongly connected components
| SCCs

\knowledge{notion}
| sink
| sinks

\knowledge{notion}
| final@SCC
| final SCC

\knowledge{notion}
| recurrent

\knowledge{notion}
| transient
| transient vertex

\knowledge{notion}
| directed acyclic graph
| DAG

\knowledge{notion}
| state-based automata
| state-based 
| State-based
| state-based@aut
| transition-based
| transition-based@aut
| transition-based automata
| state-based automaton
| over transitions
| transition-based acceptance
| transition-based automaton 
| state-based acceptance

\knowledge{notion}
| underlying graph@aut
| underlying graphs@aut

\knowledge{notion}
| initial vertex@TS
| initial vertices@TS
| initial vertex
| initial set of vertices@TS
| initial vertices

\knowledge{notion}
| initial state
| initial@state
| initial states
| initial states@aut
| initial state@aut
| initial

\knowledge{notion}
| size@aut

\knowledge{notion}
| minimal

\knowledge{notion}
| labelling with colours
| labelling
| colour labelling
| output colours@aut
| output colour@aut
| colours@aut
| couleurs@aut
| colour@aut

\knowledge{notion}
| output alphabet

\knowledge{notion}
| composition

\knowledge{notion}
| accessible@vertex
| reachable
| accessible
| unreachable

\knowledge{notion}
| accessible from
| accessible@fromVertex

\knowledge{notion}
| accessible@set
| accessible@sets

\knowledge{notion}
| run
| runs
| run on
| run over $w$
| run@automaton

\knowledge{notion}
| accepting@run
| accepting
| acceptance@run
| accepting run
| accepting runs
| acceptance@runs
| acceptance of runs
| accepting@runAut
| accepting@aut
| acceptance@runAut
| accepting run@aut
| accepting runs@aut
| acceptance@runAut
| acceptance of runs@aut
| accept@word
| rejecting@run
| rejecting run
| rejecting
| rejecting@runAut
| rejecting run@aut
| accepted@word
| accepts@aut
| accepted by@aut
| accepted
| accepted by
| accepted@aut
| rejected@aut
| accepting word

\knowledge{notion}
| vertex-labelling
| vertex-labellings
| vertex-labelled

\knowledge{notion}
| edge-labelled
| edge-labelled@graph

\knowledge{notion}
| product construction
| composition
| product automaton
| product@aut
| composing
| composition of automata
| compositions
| composition@aut
| composition operation@aut

\knowledge{notion}
| acceptance set
| acceptance sets
| acceptance set@TS
| acceptance set@TS
| acceptance sets@TS

\knowledge{notion}
| acceptance set@aut
| acceptance sets@aut

\knowledge{notion}
| acceptance condition
| acceptance conditions
| conditions
| condition
| condition@accep
| condition@acc
| conditions@acc

\knowledge{notion}
| prefix-independent
| prefix-independence

\knowledge{notion}
| automaton
| (non-deterministic) 
| automata
| (non-deterministic) automaton
| $\oo $-automata
| automata over infinite words
| NFA
| NFAs

\knowledge{notion}
| input letters
| input alphabet
| input letter

\knowledge{notion}
| deterministic
| deterministic automaton
| deterministic automata
| Deterministic
| non-deterministic automata
| non-deterministic
| non-deterministic automaton
| non-determinism
| determinism

\knowledge{notion}
| equivalent@automaton
| equivalent@automata
| equivalent@aut
| equivalence@aut
| equivalent to@aut
| equivalent to
| equivalent

\knowledge{notion}
| transition function

\knowledge{notion}
| $\SS $-complete
| complete
| completeness
| completeness@aut

\knowledge{notion}
| output of@aut
| output@aut
| output@run

\knowledge{notion}
| morphism
| morphisms 
| morphism of automaton structures

\knowledge{notion}
| automaton structure
| automaton structures

\knowledge{notion}
| residuals
| residual
| residual languages
| residual language
| residual of $L$ with respect to $u$

\knowledge{notion}
| underlying graph@aut

\knowledge{notion}
| language accepted
| accepting@automaton
| recognising
| recognise
| recognises
| recognising@automaton
| recognising@aut
| recognised
| recognises@aut
| language@aut
| language recognised
| languages recognised
| language of an automaton
| language of the automaton

\knowledge{notion}
| $\oo$-regular language
| $\oo$-regular languages
| $\oo$-regular@language
| $\oo$-regular automata
| $\oo$-regular
| $\oo$-regularity
| $\oo$-regular objective
| $\omega$-regular objectives
| $\omega$-regular languages
| non-$\oo $-regular
| $\oo $-regular conditions
| $\oo $-regular objectives

\knowledge{notion}
| $a$-self loop
| $v$-self loop
| $u$-self loop
| $v$-self loops
| self loops
| self loop

\knowledge{notion}
| history-deterministic
| HD-parity automaton
| history-determinism
| history-deterministic@aut
| HD@aut
| HD
| history-determinism@aut
| HD automaton
| HD automata
| (history-)deterministic
| history-deterministic automaton
| History-deterministic automata
| history-deterministic automata
| History-
| deterministic@hist
| History-deterministic
| déterminisme en histoire
| déterministes en histoire
| history-
| History-determinism

\knowledge{notion}
| good-for-games
| good-for-gameness
| GFG
| GFG automata

\knowledge{notion}
| pruned 
| prune

\knowledge{notion}
| Determinisable by pruning (DBP)
| determinisable by pruning

\knowledge{notion}
| resolver
| resolvers
| resolver@aut
| resolvers@aut
| resolve@aut

\knowledge{notion}
| resolver-trim

\knowledge{notion}
| sound@aut
| sound resolver@aut
| soundness@resolverAut
| sound@resolverAut
| sound resolver
| sound@resAut

\knowledge{notion}
| run induced by this resolver
| run induced by@aut
| induced by $\resolv$@aut
| induced run of $\resolv $
| run induced over $w$
| induced run
| run induced by $\ss$ over $w$
| run induced over
| induced run of $\resolv $ over $u_0$
| induced run of $\resolv $ over
| run induced by $\resolv $
| induced by@resolver
| run induced by@res
| run induced over $w$@res
| induced by@resolv
| induce@resolv
| induced by
| path induced by

\knowledge{notion}
| reachable using some sound resolver
| reachable using@resolver
| reachable using a sound resolver 
| reachable using this resolver
| reachable using $\resolv $
| reachable using the resolver $(r_0, \resolv )$
| reachable using $(r_0, \resolv )$

\knowledge{notion}
| size of the representation
| size of its representation
| representation@aut
| representation@automata

\knowledge{notion}
| safe components
| safe component

\knowledge{notion}
| safe language of a state $q$
| safe language
| safe languages

\knowledge{notion}
| safe path
| safe@path

\knowledge{notion}
| safe deterministic
| safe determinism

\knowledge{notion}
| semantically deterministic
| semantical determinism

\knowledge{notion}
| B\"uchi language
| Büchi languages
| B\"uchi@language
| B\"uchi
| Büchi
| B\"uchi languages
| (co)B\"uchi
| B\"uchi conditions
| B\"uchi condition

\knowledge{notion}
| Büchi recognisable
| B\"uchi recognisable
| coB\"uchi recognisable
| (co)B\"uchi recognisable

\knowledge{notion}
| coBüchi conditions
| coB\"uchi conditions
| coBüchi condition
| coB\"uchi language
| coB\"uchi languages
| coB\"uchi
| coB\"uchi@language
| coBüchi

\knowledge{notion}
| (co)B\"uchi automaton
| B\"uchi automaton
| Büchi automaton
| B\"uchi automata
| Büchi automata
| coB\"uchi automaton
| coBüchi automaton
| coB\"uchi automata
| coBüchi automata
| generalised B\"uchi automaton
| generalised (co)B\"uchi automaton
| generalised (co)B\"uchi automata
| generalised Büchi automaton
| generalised B\"uchi automata
| generalised Büchi automata
| generalised (co)Büchi automaton
| X-automaton
| generalised coB\"uchi automaton
| generalised coBüchi automaton
| generalised coB\"uchi automata
| generalised coBüchi automata

\knowledge{notion}
| B\"uchi transitions
| Büchi transition
| Büchi transitions

\knowledge{notion}
| coBüchi@transition
| coB\"uchi@transition
| CoB\"uchi transitions
| coB\"uchi transitions
| coB\"uchi transition
| coBüchi transition
| coBüchi transitions

\knowledge{notion}
| safe
| safety

\knowledge{notion}
| reach
| reachability

\knowledge{notion}
| coB\"uchi language associated to

\knowledge{notion}
| deterministic parity hierarchy

\knowledge{notion}
| parity index
| parity index is not less
| parity index@atMost  
| parity index at most
| parity indices

\knowledge{notion}
| letter game
| Letter game

\knowledge{notion}
| duplicated edges
| duplicated transitions

\knowledge{notion}
| language recognised by $q$
| accepted@state
| accepting@state
| recognising@state

\knowledge{notion}
| residual associated to $q$

\knowledge{notion}
| equivalent@state
| equivalence@state
| equivalent states
| equivalent@states

\knowledge{notion}
| output colour
| output colours

\knowledge{notion}
| generalised B\"uchi language
| generalised B\"uchi@language
| generalised B\"uchi
| generalised B\"uchi languages
| generalised Büchi
| generalised B\"uchi-
| gen. B\"uchi
| generalised (co)Büchi
| generalised (co)B\"uchi
| Generalised (co)B\"uchi
| Büchi généralisés
| generalised (co)B\"uchi languages
| (generalised) B\"uchi
| (generalised) coB\"uchi
| generalised (co)Büchi

\knowledge{notion}
| generalised coB\"uchi language associated to
| generalised coB\"uchi language
| generalised coB\"uchi
| coB\"uchi@generalised
| gen. coB\"uchi
| generalised coBüchi
| Generalised coB\"uchi
| Generalised coBüchi
| Generalised coB\"uchi language
| Generalised coB\"uchi
| (generalised) coB\"uchi

\knowledge{notion}
| recolour
| recolouring
| recoloured

\knowledge{notion}
| $(\S ,m)$-cover

\knowledge{notion}
| cover
| vertex cover

\knowledge{notion}
| colouring@chrm
| colouring@graph
| $k$-colouring@graph
| 3-colouring
| $k$-colouring

\knowledge{notion}
| $|V|$-colourable
| $k$-colourable
| 3-colourable

\knowledge{notion}  
| 3-colouring problem
| colouring problem

\knowledge{notion}  
| size@colouring

\knowledge{notion}
| chromatic number

\knowledge{notion}
| independent set
| independent sets

\knowledge{notion} 
| clique
| $k$-clique
| 3-clique
| 4-clique

\knowledge{notion}
| triangle-full

\knowledge{notion}
| 4-clique-free

\knowledge{notion}
| $v$-cycle
| $v$-cycles
| $t$-cycle
| $u$-cycle

\knowledge{notion}
| $\neighbourhoodOpen{v}$-cycle
| $\neighbourhoodOpen{u}$-cycle
| $\neighbourhoodOpen{v}$-cycles

\knowledge{notion}
| $G_2$
| $G_2$ game

\knowledge{notion}
| $G_2$ conjecture

\knowledge{notion}
| safe centralised
| safe centrality

\knowledge{notion}
| safe minimal
| safe minimality
| Safe minimality

\knowledge{notion}
| nice
| nice automaton

\knowledge{notion}
| normal form

\knowledge{notion}
| local alphabet at $R$
| local alphabet at

\knowledge{notion}
| transient@res
| transient residual

\knowledge{notion}
| recurrent@res
| recurrent residual

\knowledge{notion}
| localisation of $L$ to

\knowledge{notion}
| pseudo-path
| pseudo-paths

\knowledge{notion}
| initial@pseudo

\knowledge{notion}
| stabilises around
| stabilise around
| stabilising around

\input{colours.sty}

\begin{document}

\maketitle              % typeset the header of the contribution
\begin{abstract}
%We study the problem of the minimisation of (transition-based) omega-automata using generalised B\"uchi and coB\"uchi acceptance conditions, both for deterministic and history-deterministic models. 
We present a polynomial-time algorithm minimising the number of states of history-deterministic generalised coB\"uchi automata, building on the work of Abu Radi and Kupferman on coB\"uchi automata.
On the other hand, we establish that the minimisation problem for both deterministic and history-deterministic generalised B\"uchi automata is NP-complete, as well as the problem of minimising at the same time the number of states and colours of history-deterministic generalised coB\"uchi automata.
\end{abstract}

\vskip0.5em
This document contains hyperlinks.
%Each occurrence of a "notion" is linked to its ""definition"".
On an electronic device, the reader can click on words or symbols (or just hover over them on some PDF readers) to see their definition.

\shortlong{A long version with appendices containing extra material is available at \url{https://arxiv.org/abs/2407.18090}}{}
\vskip0.5em

\section{Introduction}
\label{sec:introduction}
%\subsection{Context}

%\subparagraph{Automata over infinite words for synthesis.}
First introduced by B\"uchi to obtain the decidability of monadic second order logic over $(\NN, \mathsf{succ})$~\cite{Buchi1962decision}, automata over infinite words (also called $\oo$-automata) have become a well-established area of study in Theoretical Computer Science. Part of its success is due to its applications to model checking (verify whether a system satisfies some given specifications)~\cite{BK08PrinciplesMC,VardiWolper94ReasoningInf,EKV21Verification} and synthesis (given a set of specifications, automatically construct a system satisfying them)~\cite{BL69Strategies,PR89Synthesis}.
In many of these applications, mainly in problems related to synthesis, "non-deterministic" models of automata are not well-suited, and costly determinisation procedures are usually needed~\cite{Safra1988onthecomplexity}.

%\subparagraph{History-determinism.}
In 2006, Henzinger and Piterman~\cite{HP06} proposed\footnote{Similar ideas had been previously investigated by Kupferman, Safra and Vardi~\cite{KSV96Relating}, and Colcombet studied "history-determinism" in the context of cost functions~\cite{Colcombet2009CostFunctions}.} a model of automata, called \emph{"history-deterministic"} ("HD")\footnote{These automata were first introduced under the name \emph{good-for-games} (GFG). Currently, these two notions are no longer used interchangeably, although they coincide in the case of $\oo$-automata. We refer to the survey~\cite{BL23SurveyHD} for further discussions.}, presenting a restricted amount of "non-determinism" so that they exactly satisfy the properties that are needed for applications in synthesis.
Namely, these automata do not need to guess the future: an automaton is "history-deterministic" if it admits a strategy resolving the non-determinism on the fly, in such a way that the "run" built by the strategy is "accepting@@run" whenever the input word belongs to the "language of the automaton".
Since their introduction, several lines of research have focused on questions such as the succinctness of "history-deterministic" automata~\cite{KS15DeterminisationGFG,CCL22SizeGFG}, the problem of recognising them~\cite{BK18,BKLS20SuccRec}, or extensions to other settings~\cite{LZ20,BL19GFGFromND,BL21HDvsGFG}.

%Kuperberg and Skrzypczak showed that, for some classes of $\oo$-automata, "history-deterministic" automata might be exponentially smaller than "equivalent@@aut" "deterministic" ones~\cite{KS15DeterminisationGFG}.

%\subparagraph{The minimisation problem: state-of-the-art.}
Minimisation of automata stands as one of the most fundamental problems in automata theory, for various reasons. Firstly, for its applications: when employing algorithms that rely on automata, having the smallest possible ones is crucial for efficiency.
Secondly, beneath the problem of minimisation lies a profoundly fundamental question: What is the essential information needed to represent a formal language?
A cornerstone result about automata over finite words is that each regular language admits a unique minimal deterministic automaton, in which states corresponds to the "residuals" of the language (the equivalence classes of the Myhill-Nerode congruence).
Moreover, this minimal automaton can be obtained from an equivalent deterministic automaton with $n$ states in time $\O(n\log n)$~\cite{Hopcroft71nLognAlgorithm}.
%in time $\O(n\log n)$, where $n$ is the size of an input "deterministic" automaton~\cite{Hopcroft71nLognAlgorithm}.

However, the situation is quite different in the case of $\oo$-automata. Contrary to the case of finite words, the "residuals" of a language are not sufficient to construct a correct "deterministic" automaton in general.
In 2010, Schewe proved that the minimisation of "deterministic" "B\"uchi" "automata" is $\NP$-complete~\cite{Schewe10MinimisingNPComplete}. That appeared to be a conclusion to the minimisation problem, but a crucial aspect of his proof was that the $\NP$-completeness is established for automata with the acceptance condition \emph{over states}, and this proof does not generalise to \emph{transition-based} automata.
A surprising positive result was obtained in 2019 by Abu Radi and Kupferman: we can minimise  "history-deterministic" "coB\"uchi" automata using transition-based acceptance in polynomial time~\cite{AK19Min}.
One year later, Schewe showed that the very same problem becomes $\NP$-complete if state-based acceptance is used~\cite{Schewe20MinimisingGFG}.
Multiple other results have backed the idea that transition-based acceptance is a better-suited model; we refer to~\cite[Chapter~VI]{Casares23PhD} for a detailed discussion.
The work of Abu Radi and Kupferman raised the question of what is the complexity of the minimisation problem for other classes of transition-based automata such as "(history-)deterministic" "B\"uchi" automata.
Since then, to the best of our knowledge, the only further result concerning minimisation of transition-based automata is Casares' $\NP$-completeness proof for the problem of minimising "deterministic" Rabin automata~\cite{Casares2021Chromatic}.

%\subparagraph{Generalised (co)B\"uchi automata.}
In this paper, we focus our attention on "generalised B\"uchi" and "generalised coB\"uchi" automata, in which the "acceptance condition" is given, respectively, by conjunctions of clauses ``see colour $c$ infinitely often'', and by disjunctions of ``eventually avoid colour $d$''.
"Generalised (co)B\"uchi" automata are as expressive as (co)B\"uchi automata, but they can be more succinct, due to their more complex "acceptance condition".
These automata appear naturally in the model-checking and synthesis of temporal properties~\cite{CVWY92Memory,GPVW95OnTheFly,SomenziBloem00EfficientBuchi}; for instance, \texttt{SPOT}'s LTL-synthesis tool transforms a given LTL formula into a "generalised B\"uchi" automaton~\cite{DLFMRX16Spot2,MichaudColange18Synt}. Also, many efficient algorithms for their emptiness check have been developed~\cite{RDKP13SCCEmpt,RDKP17EmptinessGenB,BDS20Seminator2}.

Several works have approached the problem of reducing the state-space of "generalised B\"uchi" automata, which is usually done either by the use of simulations~\cite{SomenziBloem00EfficientBuchi,JuvekarPiterman2006} (which do not yield minimal automata), or by the application of SAT solvers~\cite{Ehlers10MinimisingSAT, BD14Mechanizing}. However, to the best of our knowledge, no theoretical result about the exact complexity of this minimisation problem appears in the literature.

\subsection*{Contributions}
%We outline the main contributions of the paper:
We provide a polynomial-time minimisation algorithm for "history-deterministic" "generalised coB\"uchi" automata (Theorem~\ref{th:minimisationHDgenCoBuchi}). %Our construction uses  Abu Radi-Kupferman's minimal "history-deterministic" "coB\"uchi" automaton as starting point, and we show how we can reduce its state-space in an optimal way by using a "generalised coB\"uchi" condition.
Our algorithm uses  Abu Radi-Kupferman's minimal "history-deterministic" "coB\"uchi" automaton as a starting point, and reduces the state-space of this automaton in an optimal way to use a "generalised coB\"uchi" condition.

We prove that the minimisation problem is $\NP$-complete for "history-deterministic" "generalised B\"uchi" automata (Theorem~\ref{thm-NPhard:minHDBuchi}), as well as for "deterministic" "generalised B\"uchi" and "generalised coB\"uchi" automata (Theorem~\ref{thm-NPhard:mindetcoBuchi}). We remark that both the $\NP$-hardness and the $\NP$-upper bound are challenging. Indeed, to obtain that the problem is in $\NP$, we first need to prove that a minimal "HD" "generalised B\"uchi" automaton only uses a polynomial number of "output colours". Additionally, we adapt a proof from~\cite{CM24Simplifying} to show that minimising at the same time the number of states and colours is $\NP$-complete for all the previous models, including "history-deterministic" "generalised coB\"uchi" automata (Theorem~\ref{thm-NPHard:min-state-and-col-NPhard}).

We summarise the results about the state-minimisation of transition-based automata in Table~\ref{table:complexity}.

\setlength\extrarowheight{7pt}
\newcolumntype{M}[1]{>{\centering\arraybackslash}m{#1}} %Center content and set a fixed width. Also, replace p with m
\newcolumntype{N}{@{}m{0pt}@{}}
%\renewcommand{\thefootnote}{\fnsymbol{footnote}}
%\newcommand{\mycolwidth}{1.3cm}
%\hdashline[10pt/2pt] %Instead of \hline, if dashed wanted

\begin{table}[h]
	\centering
	\begin{tabular}{|M{3.5cm}||M{1.8cm}|M{1.8cm}|M{2.3cm}|M{2.3cm}|N} 
		\hline
		\diagbox[innerleftsep=3mm,innerrightsep=-2mm, outerrightsep=2mm]{\textbf{Model}}{\textbf{Condition}} & "coB\"uchi"   &   "B\"uchi"    & "generalised coB\"uchi"     & "generalised B\"uchi"              &\\[0mm]
		\hline
		\hline

		"Deterministic" & Unknown  & Unknown    & $\NPc$ (Theorem~\ref{thm-NPhard:mindetcoBuchi})     & $\NPc$ (Theorem~\ref{thm-NPhard:mindetcoBuchi})  & \\[4mm]
		\hline
		
		"History-deterministic" & $\PTimeFull$~\cite{AK22MinimizingGFG}  & Unknown    & $\PTimeFull$ (Theorem~\ref{th:minimisationHDgenCoBuchi})     & $\NPc$ (Theorem~\ref{thm-NPhard:minHDBuchi})  & \\[4mm]
		\hline
	\end{tabular}
			\vspace{1mm}
	\caption{Complexity of the minimisation problem for different types of transition-based automata.}
	\label{table:complexity}
\end{table}

We note that the $\PTimeFull$ complexity of recognising HD automata can be lifted from Büchi and coBüchi conditions to their generalised versions (\Cref{cor:HDness}).
This result can be considered folklore, although we have not find it explicitly in the literature.
In \shortapp{app:G2-conjecture}, we also lift the  characterisation based on the "$G_2$ game" from (co)Büchi automata to generalised ones \shortlong{}{(\Cref{th-FP:G2-conj-genBuchi})} (a similar remark was suggested in the conclusion of~\cite{BKLS20SuccRec}).

\subparagraph{State-minimality.} In this paper, we primarily focus our attention on the minimisation of the number of \emph{states} of the automata.
In Theorem~\ref{thm-NPHard:min-state-and-col-NPhard} we also consider the minimisation of both the number of states and colours of the "acceptance condition".
We highlight that the decision on how we measure the size of the automata is orthogonal to putting the acceptance condition over transitions.

The reader may wonder why we focus on these quantities and not, e.g., on the number of transitions. This choice, which is standard in the literature (\cite{AK22MinimizingGFG,Schewe10MinimisingNPComplete}), is justified by various reasons. First, the number of transitions of an automaton is polynomial in the number of states. Indeed, we can assume that there are no two transitions between two states over the same input letter (see~\cite[Prop.18]{CCL22SizeGFG}), therefore, $|\DD|\leq |Q|^2|\Sigma|$.
Maybe more importantly, in the case of automata over finite words, each state of the minimal automaton carry a precise information about the language it recognises: a residual of it. Ideally, a construction for a state-minimal automaton for an $\oo$-regular language should lead to an understanding of the essential information necessary to represent it.

The interest of minimising both the number of states and the number of colours comes from the fact that the number of colours can be exponential on the number of states, but the "size of the representation" of the automaton is polynomial in the sum of these quantities.

\section{Preliminaries}
\label{sec:preliminaries}
%%Section essentially copied from the definition of the article on explorable automata by Emile Hazard and Denis
%\subsection{Preliminaries}\label{sec:defs}
%We denote by $[i,j]$ the integer interval $\{i,i+1,\dots,j\}$ and let $[j]=[1,j]$. The cardinality of a set $S$ is written $|S|$, and its powerset $\pow{S}$. 
The disjoint union of two sets $A,B$ is written $A \disjUnion B$.
The \intro*"empty word" is denoted $\intro*\emptyword$.
A \intro*"factor" of a word $w$ is a word $u$ such that there exist words $x,y$ with $w=xuy$.
For an infinite word $w\in \SS^\oo$, we denote $\intro*\minf(w)$ the set of letters occurring infinitely often in $w$.

\subsection{Automata}
We let $\Sigma$ be a finite alphabet.
\AP An ""automaton"" is a tuple $\A = (Q,\SS, q_{\init}, \DD, \GG, \colAut, W)$, where $Q$ is its set of states, 
$q_{\init}$ its ""initial state"", 
$\Delta\subseteq Q\times\Sigma\times Q$ its set of transitions, 
$\GG$ its ""output alphabet"", 
$\colAut\colon \DD \to \GG$ a ""labelling with colours"", and $W\subseteq \GG ^\oo$ its ""acceptance condition"".
%For $(p,a)\in Q\times\Sigma$, we will note  $\Delta(p,a)=\{q\in Q, (p,a,q)\in\Delta\}$. If $X\subseteq Q$, we note $\Delta(X,a)=\bigcup_{p\in X} \Delta(p,a)$.
\AP A state $q$ is called ""reachable"" if there exists a path from $q_\init$ to $q$.
\AP The ""size@@aut"" of an "automaton" is its number of states, written $|Q|$.
\AP% Unless stated otherwise, a ("deterministic"/"history-deterministic") automaton is said ""minimal"" if it has a minimal number of states amongst "equivalent@@aut" ("deterministic"/"history-deterministic") automata.
We write $p\re{a:c}q$ if $(p,a,q)\in \DD$ and $\colAut((p,a,q)) = c$. 

\AP A ""run"" $\rr$ on an infinite word $w=a_1 a_2\dots\in\Sigma^\omega$ is an infinite sequence of transitions $\rho = (q_{0},a_1,q_1)(q_{1},a_2,q_2)(q_{2},a_3,q_3),\dots\in\Delta^\omega$ with $q_0 = q_\init$.
\AP It is ""accepting@@run"" if the infinite word 
%$\colAut((q_{0},a_1,q_1)), \colAut(q_1, a_2, q_2), \colAut(q_2, a_3, q_3) \cdots \in \GG^\omega$,
$c_1c_2\dots\in \GG^\oo$ defined by $c_i = \colAut(q_{i-1},a_i,q_i)$, called the ""output@@run"" of $\rho$, belongs to $W$.
%This infinite word is sometimes denoted $\intro*\outRun(\rr)$. 

\AP The \intro*"language of an automaton" $\A$, denoted $\intro*\Lang{\A}$, is the set of words that admit an "accepting run".
\AP We say that two automata $\A$ and $\B$ over the same alphabet are ""equivalent@@aut"" if $\Lang{\A}=\Lang{\B}$.

\AP An "automaton" $\A$ is ""deterministic"" (resp. ""complete"") if, for all $(p,a) \in Q \times \SS$, there exists at most (resp. at least) one $q \in Q$ such that $(p,a,q) \in \DD$. We note that if $\A$ is "deterministic", a word $w \in \SS ^\oo$ admits at most one "run" in $\A$.

\subsection{Acceptance conditions}

In this paper we will focus on automata using "generalised B\"uchi" and "generalised coB\"uchi" acceptance conditions. A "generalised B\"uchi" condition can be seen as a conjunction of "B\"uchi conditions", while a "generalised coB\"uchi" condition can be seen as a disjunction of "coB\"uchi conditions".

\AP A ""generalised B\"uchi"" condition with $k$ colours is defined over the "output alphabet" $\GG = \pow{C}$, with $C$ a set of $k$ ""output colours"", as
\[\intro*\genBuchiCol{C} = \{w\in \GG^\oo \mid \bigcup \minf(w) = C\}.\]
It contains sequences of sets of colours such that every colour is seen infinitely often.
Usually, we take $C= [k] = \{1,2,\dots,k\}$.

\AP The dual condition is the ""generalised coB\"uchi"" condition with $k$ colours. That is, we define: 
\[\intro*\genCoBuchiCol{C} = \{w\in \GG^\oo \mid \bigcup \minf(w) \neq C\}.\]
It contains sequences of sets of colours such that at least one colour is seen finitely often.

%In general, we will assume that $\GG= \pow{\{1,\dots,k\}}$ and that $A_i = \{i\}$, and we will simply write $\genBuchiCol{\GG}$ and $\genCoBuchi_\GG$.\\
\AP The ""size of the representation"" of an automaton using a "generalised (co)B\"uchi" condition  with $k$ colours is $|Q| + |\SS| + k$; such an automaton can be described in polynomial space in this measure.

\AP A ""B\"uchi condition"" (resp. ""coBüchi condition"") can be defined as a "generalised B\"uchi" (resp. "generalised coB\"uchi") condition in which $k=1$.
In this case, we call ""B\"uchi transitions"" (resp. ""coB\"uchi transitions"") the transitions $(p,a,q)\in \DD$ such that $\colAut((p,a,q)) = \set{1}$.
\AP An "automaton" using an "acceptance condition" of type $X$ is called an ""X-automaton"".

\AP A language $L \subseteq \SS^\oo$ is ""(co)B\"uchi recognisable"" if there exists a "deterministic" "(co)B\"uchi automaton" $\A$ such that $\A$ "recognises" $L$.
These coincide with languages "recognised" by "generalised (co)B\"uchi" automata (see \Cref{cor-first:gen-Buchi-to-Buchi}).
We note that non-deterministic "(generalised) B\"uchi" automata are strictly more expressive, while "non-deterministic" "(generalised) coB\"uchi" automata are as expressive as "deterministic" ones~\cite{MiyanoH84Alternating}.

\subsection{History-determinism}
\AP An automaton is called "history-deterministic" (or "HD" for short), if there exists a function, called a "resolver", that resolves the non-determinism of $\A$ depending only on the prefix of the input word read so far. Formally, a  ""resolver"" for an automaton $\A$ is a function $\sigma : \Sigma^+ \rightarrow \DD$ such that for all words $w = a_0a_1\dots \in \SS^\oo$, the sequence $\intro*\resolvPath{\ss}(w) = \sigma(a_0)\sigma(a_0a_1)\sigma(a_0a_1a_2)\dots \in \DD^\oo$ (\AP called the ""run induced by $\ss$ over $w$"") satisfies:
\begin{enumerate}
	\item $\resolvPath{\ss}(w)$ is a "run on" $w$ in $\A$,
	\item if $w\in \Lang{\A}$, then $\resolvPath{\ss}(w)$ is an "accepting run".
\end{enumerate}

\AP An automaton is \intro*"history-deterministic" if it admits a "resolver".
\AP We say that a ("deterministic"/"history-deterministic") automaton is  ""minimal"" if it has a minimal number of states amongst "equivalent@@aut" ("deterministic"/"history-deterministic") automata.

%\AP An automaton is \intro*"history-deterministic" (or "HD" for short), if there exists a function $\sigma : \Sigma^+ \rightarrow \DD$, \AP called a ""resolver"", that resolves the non-determinism of $\A$ depending only on the prefix of the input $\omega$-word read so far: over every $\omega$-word $w$, the function $n \mapsto \sigma(w_0 w_1 \dots w_{n-1})$ induces a sequence that is a run of $\A$ over $w$ (\AP called the ""run induced by $\ss$ over $w$""), and it is "accepting@@run" whenever $w \in \Lang{\A}$. 
%\AP We write $\intro*\resolvPath{\ss}: \Sigma^+ \rightarrow Q$ for the function associating each finite word $w$ to the state reached by following the "run induced over" $w$ by the "resolver" $\sigma$.

\begin{remark}
	Every deterministic automaton is "HD".
	While the converse is false (see Example~\ref{ex:fin-b-or-fin-c} below), we note that any language $L\subseteq \SS^\oo$ recognised by an "HD" "B\"uchi automaton" (resp. "coB\"uchi automaton"), can be "recognised" by a "deterministic" "B\"uchi automaton" (resp. "deterministic" "coB\"uchi automaton")~\cite{KSV96Relating}.
\end{remark}

\begin{figure}[H]
	\centering
	\includegraphics[]{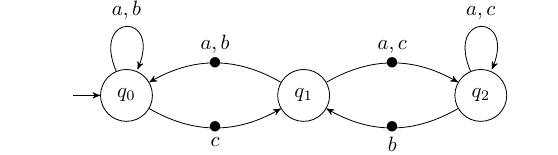}
	\caption{A "history-deterministic" "coB\"uchi" automaton "recognising" the language $ L = \{w\in \SS^\oo \mid \{b,c\} \nsubseteq \minf(w)\}$.
		"CoB\"uchi transitions" are represented with a dot on them.
	}       
	\label{fig-FP:coBuchi-b-or-c}
\end{figure}

\begin{example}[{From~\cite[Ex.~2.3]{CCFL24FromMtoP}}]\label{ex:fin-b-or-fin-c}
	Let $\SS = \{a,b,c\}$ and
	$ L = \{w\in \SS^\oo \mid \{b,c\} \nsubseteq \minf(w)\}$,
	that is, $L$ is the set of words that contain either $b$ or $c$ only finitely often.
	
	In Figure~\ref{fig-FP:coBuchi-b-or-c} we show an automaton "recognising" $L$ that is not \emph{determinisable by prunning}, that is, it cannot be made "deterministic" just by removing transitions.
	
	We claim that this automaton is "history-deterministic". First, we remark that the only non-deterministic choice appears when reading letter $a$ from the state $q_1$. A "resolver" can be defined as follows: whenever we have arrived at $q_1$ from $q_0$ (by reading letter $c$), if we are given letter $a$ we go to state $q_2$; if we have arrived at $q_1$ from $q_2$ (by reading letter $b$), we will go to state $q_0$.
	Therefore, if after some point letter $b$  (resp. letter $c$) does not appear, we will stay forever in state $q_2$ (resp. state $q_1$) and accept.	
\end{example}

\subsection{Residuals and prefix-independence}
Let $L\subseteq \SS^\oo$ and $u\in \SS^*$. \AP The ""residual of $L$ with respect to $u$"" is the language
\[ \intro*{\quot{u}{L}} = \{w\in \SS^\oo \mid uw\in L \}. \] 
\AP We write $\intro*{\resClassLang{u}} = \{v\in \SS^* \mid \quot{u}{L} = \quot{v}{L}\}$, and $\intro*\Res{L}$ for the set of "residuals" of a language~$L$.

\AP Given an "automaton" $\A$ and a state $q$, we denote $\intro*\initialAut{\A}{q}$ the automaton obtained by setting $q$ as "initial state", and we refer to $\Lang{\initialAut{\A}{q}}$ as the ""language recognised by $q$"".
We say that two states $q,p$ are ""equivalent@@state"", written $q \intro*\eqRes p$, if they recognise the same language.
We note $\intro*{\resClassAut{q}{\A}}$ the set of states "equivalent@@state" to $q$ (we simply write $\intro*{\resClass{q}}$ when $\A$ is clear from the context).

\AP We say that an "automaton" $\A$ is ""semantically deterministic"" if non-deterministic choices lead to equivalent states, that is, if for every state $q$ and pair of transitions $q\re{a} p_1$, $q\re{a}p_2$ we have $p_1\eqRes p_2$.

If $\A$ is "semantically deterministic" and $u\in \SS^*$ is a word labelling a path from the initial state to $q$, then $\Lang{\initialAut{\A}{q}} = \quot{u}{L}$.
\AP We say then that $\quot{u}{L}$ is the ""residual associated to $q$"".
%In this case, we identify $\resClassLang{u}$ with $\resClassAut{q}{\A}$, and allow notation such as $v\in \resClass{q}$, if $v$ is a word labelling a path from the initial state to $q$.
\AP For a "residual" $R\in \Res{L}$ we denote $\intro*\resStates{R}$ the set of states of $\A$ "recognising@@state" $R$. We remark that $\resStates{R} = \resClassAut{q}{\A}$ for any state $q$ "recognising@@state" $R$.

\AP We say that $L$ is ""prefix-independent"" if for all $w\in \SS^\oo$ and $u\in \SS^*$, $w\in L \iff uw \in L$.

\begin{remark}
	A language $L$ is "prefix-independent" if and only if it has a single "residual". 
\end{remark}

\subsection{Morphisms of automaton structures}
\AP An ""automaton structure"" over an alphabet $\SS$ is a tuple $\S=(Q,\DD)$, where $\DD\subseteq Q\times \SS \times Q$. 
Let $\S_1 = (Q_1,\DD_1)$ and $\S_2=(Q_2,\DD_2)$ be two "automaton structures" over the same alphabet. 
\AP A ""morphism of automaton structures"" is a mappings $\phi\colon Q_1\to Q_2$ such that for every $(q,a,q')\in \DD_1$,  $(\phi(q),a,\phi(q'))\in \DD_2$.
We note that such a "morphism" induces a function $\phi_\DD \colon \DD_1\to \DD_2$ sending $(q,a,q')$ to $(\phi(q),a,\phi(q'))$. We also denote this function $\phi$, whenever no confusion arises, and denote a "morphism" of "automaton structures" by $\phi\colon \S_1\to \S_2$.

\section{First properties and examples}
We discuss a further example of a "history-deterministic" automaton and state some well-known facts about these automata that will be relevant for the rest of the paper.

\subsection{A central example}

The following automata will be used as a running example in Section~\ref{sec:polynomial-HD-coBuchi}.

\begin{figure}[ht]
	%\centering
	\hspace{0mm}
	\begin{minipage}[c]{0.45\textwidth} 
		\includegraphics[scale=1.1]{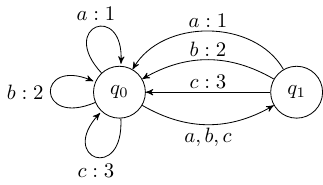}
	\end{minipage} 
	\hspace{0mm}
	\begin{minipage}[c]{0.45\textwidth} 
		\includegraphics[scale=0.9]{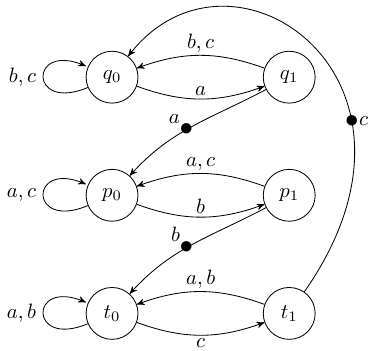}
	\end{minipage} 
	\caption{On the left, a "history-deterministic" "generalised coB\"uchi" automaton "recognising" the language $L_3$ of words eventually avoiding factor $xx$ for some letter $x$. On the right, a "minimal" "deterministic" "coB\"uchi" automaton for the same language ("coB\"uchi transitions" have a dot on them). In both cases, the initial state is irrelevant, as the language is "prefix-independent".}
	\label{fig-FP:aut-factor-xx}
\end{figure}

\begin{example}\label{ex:fin-xx} \ap{Reviewer 3 asked us to explain the connection with the one in ICALP 2013 paper. I see that the languages are related, but the examples themselves aren't. Thoughts?} Let $\SS_n$ be an alphabet of size $n$, and let \[ L_n = \{w\in \SS_n^\oo \mid \text{ for some } x\in \SS_n \text{ the factor } xx \text{ appears only finitely often in } w\}. \]
	On the left of Figure~\ref{fig-FP:aut-factor-xx} we show a "history-deterministic" "generalised coB\"uchi" automaton "recognising" $L_n$ with just $2$ states (we show it for $\SS_3= \{a,b,c\}$, but the construction clearly generalises to any $n$).
	The set of colours is $C = \{1,2,3\}$, and we accept if eventually some colour is not produced. 
	A "resolver" can be defined as follows: in a round-robin fashion, we bet that the factor that does not appear is $aa$, then $bb$, then $cc$. While factor $aa$ is not seen, we will take transition $q_0 \re{a} q_1$ whenever letter $a$ is read, to try to avoid colour $1$. Whenever factor $aa$ is read, we switch to the corresponding strategy with letter $b$, trying to avoid colour $2$. If eventually factor $xx$ is not produced, for $x\in \{a,b,c\}$, then some colour will forever be avoided. 
	
	We show a "deterministic" "coB\"uchi" automaton for $L_3$ on the right of Figure~\ref{fig-FP:aut-factor-xx}.
	Applying the characterisation of Abu Radi and Kupferman~\cite{AK22MinimizingGFG} (see Lemma~\ref{lem:minimal-coBuchi-AK}), we can prove that this automaton is "minimal" amongst "HD" "coB\"uchi" automata.
	More generally, we can prove that a "minimal" "HD" "coB\"uchi" automaton for $L_n$ has at least $2n$ states , and in fact, in this case, this optimal bound can be achieved with a "deterministic" automaton.	
\end{example}

\subsection{Duality B\"uchi - coB\"uchi}

\begin{remark}\label{rmk-FP:duality-Buchi-coBuchi}
	Let $\A$ be a "deterministic" "generalised B\"uchi" "automaton" of size $n$ and using $k$ "output colours". 
	It suffices to replace the "acceptance condition" $\genBuchiCol{[k]}$ with $\genCoBuchiCol{[k]}$ to obtain a "deterministic" "generalised coB\"uchi" "automaton" of size $n$ and using $k$ output colours "recognising" the complement language $\SS^\oo\setminus \Lang{\A}$.
	Symmetrically, we can turn any "deterministic" "generalised coB\"uchi" "automaton" into a "deterministic" "generalised B\"uchi" "automaton" with the same number of states and colours recognising the complement language.	
	As a consequence, the minimisations of "deterministic" "generalised B\"uchi" "automaton" and "deterministic" "generalised coB\"uchi" "automaton" are linear-time-equivalent problems.
\end{remark}

We highlight that the hypothesis of "determinism" in the previous remark is crucial. This duality property no longer holds for "non-deterministic" (or "history-deterministic") automata.

\begin{lemma}%[\cite{KS15DeterminisationGFG}]
	There exists a "history-deterministic" "generalised coB\"uchi" automaton $\A$ such that any "history-deterministic" "generalised B\"uchi" automaton "recognising"  $\SS^\oo\setminus \Lang{\A}$ has strictly more states than $\A$.
\end{lemma}

Such an example is provided by the language $L_3$ from Example~\ref{ex:fin-xx} (\shortlong{in the long version we }{in Lemma~\ref{lem-NPhard:aut-to-colour} we will }prove that any "non-deterministic" "generalised B\"uchi" automaton "recognising"  $\SS^\oo\setminus \Lang{\A}$  has at least $3$ states).
% (as this language coincides with $\langGraph$ for $G$ a "3-clique").
Relatedly, for the non-generalised conditions, Kuperberg and Skrzypczak showed that the gap between an "HD" "coB\"uchi" and an "HD" "B\"uchi" automaton for the complement language can be exponential as well~\cite{KS15DeterminisationGFG}, using the link between complementation and determinisation of HD automata from \cite[Thm 4]{BKKS13HDneqDBP}.

\subsection{From generalised (co)B\"uchi to (co)B\"uchi}\label{subsec-first:composition}

The idea of this construction is to define a particular deterministic coB\"uchi automaton recognizing $\genCoBuchiCol{C}$, and to associate it to the input generalised coB\"uchi automaton via a cascade composition (defined below) in order to obtain the wanted coB\"uchi automaton. We will now detail these different steps.

\subparagraph*{Deterministic coB\"uchi automaton for $\genCoBuchiCol{C}$.}

Let $C=\{1,2,\cdots,k\}$ be a set of $k$ colours; for convenience, in the context of colours, we will use the symbol $+$ to denote addition modulo $k$, in particular, $k+1 = 1$.
We build a "deterministic" "coB\"uchi" automaton $\intro*\autGenCoB{C}$ over the alphabet $\GG=\pow{C}$ "recognising" the language $\genCoBuchiCol{C}$.
It has as a state $q_i$ for each colour $i \in C$ and contains the transitions $q_i \xrightarrow{X:\emptyset} q_i$, if $i \notin X$, and $q_i \xrightarrow{X:1} q_{i+1}$, if $i \in X$, for all $X\in\Gamma$.  
The "initial state" is arbitrary.

We claim that the automaton $\autGenCoB{C}$ "recognises" the language $\genCoBuchiCol{C}$.
First, we remark that the "accepting runs" of $\autGenCoB{C}$ are exactly those that eventually remain forever in a state $q_i$.
Let $w = w_1 w_2 \dots\in \pow{C}$.
If $w$ is "accepted by" $\autGenCoB{C}$, then the "run on" $w$ eventually stays in a $q_i$, so $w$ eventually does not contain colour $i$, and $w\in \genCoBuchiCol{C}$.
Conversely, if $w$ is rejected by $\autGenCoB{C}$, it takes all transitions $q_i \xrightarrow{X:1} q_{i+1}$ infinitely often, so $w$ must contain all colours in $C$ infinitely often.

We define in a similar fashion a "deterministic" "B\"uchi automaton" $\intro*\autGenB{C}$ "recognising" the language $\genBuchiCol{C}$, simply by changing the "acceptance condition" of $\genCoBuchiCol{C}$ to $\genBuchiCol{C}$.

\begin{remark}
	The automaton $\autGenCoB{C}$ has $k$ states, but exponentially many transitions. This is made possible by the fact that its alphabet $\Gamma$ is exponential in the number of colours $k$.
\end{remark}

\subparagraph*{Cascade composition of automata.}
Let $\A = (Q_\A, \SS_\A, q_{\init}^\A, \DD_\A, \GG_\A, \colAut_\A, W_\A)$ and $\B = (Q_\B,\ab \SS_\B,\ab q_{\init}^\B,\ab \DD_\B,\ab \GG_\B, \colAut_B, W_B)$ be two automata such that $\SS_\B = \GG_\A$ (i.e., $\B$ is an automaton over the set of "output colours" of $\A$).
\AP The ""cascade composition"" of $\A$ and $\B$ is the automaton over $\SS_\A$ defined as:
\[\B \intro*\autComposition \A = (Q_\A \times Q_\B, \SS_\A, (q_{\init}^\A, q_{\init}^\B), \DD', \GG_\B, \colAut', W_\B),\]
with transitions
$(p_\A,p_\B) \re{a:c} (q_\A,q_\B)$ if $p_\A \re{a:b}q_\A\in \DD_\A$ and $p_\B \re{b:c}q_\B\in \DD_\B$.

Intuitively, given a word in $\SS_\A^\oo$, we feed the output of $\colAut_\A$ directly as input to $\B$, while keeping track of the progression in both automata. We "accept@@word" according to the "acceptance condition" of $\B$.

\begin{lemma}[Folklore]\label{lem-first:composition-preserves-hd}
	Let $\A$ be an automaton with "acceptance condition" $W\subseteq\GG^\oo$, and let $\B$ be a "deterministic" automaton over $\GG$ "recognising" $W$. 
	Then $\B \autComposition \A$ "recognises" $\Lang{\A}$, and the automaton $\B \autComposition \A$ is "history-deterministic" (resp. "deterministic") if and only if $\A$ is.	
\end{lemma}
%\begin{proof}
%	\begin{description}
%		\item[$\Lang{\B \autComposition \A}= \Lang{\A}$.] If a word $w$ is accepted in $\B \autComposition \A$ via a run $\rho$, then the projection of $\rho$ on the $\A$-component yields a run $\rho_A$ which is "accepting" (as its output is "accepted by" $\B$). Conversely, if $\rho_\A$ is an "accepting run" on $w$ in $\A$ producing $\tilde{w}$ as output, then there is a run $\rho_\B$ in $\B$ "accepting@@word" $\tilde{w}$.
%		The run on $w$ in $\B \autComposition \A$ that builds in parallel $\rho_\A$ and $\rho_\B$ is then "accepting@@run". 
%		
%		\item[History-determinism is preserved.] A "resolver" $\ss_A$ for $\A$ naturally extends to a "resolver" $\ss'$ for $\B\autComposition \A$, as the transitions in the $\B$-component are already determined, since $\B$ is assumed "deterministic". 
%		Conversely, a "resolver" for $\B\autComposition \A$ extends to one for $\A$ just by forgetting the $\B$ component of the automaton.\qedhere
%	\end{description}
%\end{proof}

Thus, to convert a "generalised coB\"uchi" automaton $\A$ to an "equivalent@@aut" "coB\"uchi" one, we can just compose it with $\autGenCoB{C}$. The symmetric result holds for "generalised B\"uchi" automata.
% to obtain the coB\"uchi automaton $\A=\D \autComposition \H$ recognising the same language as $\H$. Furthermore, it follows from \cref{lem-first:composition-preserves-hd} below that $\A$ is history-deterministic if and only if $\D$ is. 

\begin{corollary}[Folklore]\label{cor-first:gen-Buchi-to-Buchi}
	Let $\A$ be a "generalised coB\"uchi" automaton using $C$ as "output colours".
	The automaton $\autGenCoB{C} \autComposition \A$ is a "coB\"uchi" automaton  "equivalent" to $\A$ which is "(history-)deterministic" if and only if $\A$ is.
	Moreover, $\autGenCoB{C} \autComposition \A$ can be computed in polynomial time in the "size of the representation" of $\A$. The same is true for "generalised B\"uchi" automata.
\end{corollary}

%\begin{remark}[Size and computation of $\autGenCoB{C} \autComposition \A$.]
	We note that $\autGenCoB{C} \autComposition \A$ has $k\cdot|\A|$ states, where $k=|C|$, but it  might have exponentially many transitions in $k$.
	However, we underline that we can compute $\autGenCoB{C} \autComposition \A$  from $\A$ in polynomial time in the "size of its representation". 
	Indeed, we just need to compose $\A$ with the restriction of $\autGenCoB{C}$ to transitions whose letters are subsets of colours that appear in $\A$. 
%\end{remark}

\subparagraph*{Deciding history-determinism.} 
The problem of deciding whether an automaton is "HD" is known to be in $\PTimeFull$ for "B\"uchi" and "coB\"uchi" automata~\cite{KS15DeterminisationGFG,BK18,BKLS20SuccRec}.
Combining this fact with \Cref{cor-first:gen-Buchi-to-Buchi}, we directly obtain:

\begin{corollary}\label{cor:HDness}
Given a "generalised Büchi" (resp. "generalised coBüchi") "automaton", it is in $\PTimeFull$ to decide whether it is "history-deterministic".
\end{corollary}

A different proof of Corollary~\ref{cor:HDness}, which goes through the "$G_2$ game"~\cite{BK18}, is presented in \shortapp{app:G2-conjecture}.

\section{Polynomial-time minimisation of HD generalised coB\"uchi automata}
\label{sec:polynomial-HD-coBuchi}

In this section we present one of the main contributions of the paper (Theorem~\ref{th:minimisationHDgenCoBuchi}):  "history-deterministic" "generalised coB\"uchi" automata can be minimised in polynomial time.

\begin{theorem}\label{th:minimisationHDgenCoBuchi}
	Given a "history-deterministic" "generalised coB\"uchi" automaton, we can build in polynomial time in its "representation@@aut" an "equivalent@@aut" "history-deterministic" "generalised coB\"uchi" automaton with a minimal number of states.
\end{theorem}

The proof of this result strongly relies on the construction of minimal "coB\"uchi" automata by Abu Radi and Kupferman~\cite{AK22MinimizingGFG}. 
We will show that, for a "coB\"uchi recognisable" language $L$, we can extract a minimal "HD" "generalised coB\"uchi" automaton for $L$ from its minimal "HD" "coB\"uchi" automaton.

In Section~\ref{subsec-poly:coBuchi} we introduce some terminology and state the main property satisfied by the minimal "HD" "coB\"uchi" automaton of Abu Radi and Kupferman. We then present our construction, decomposing it in two steps for simplicity: first we construct a minimal "HD" "generalised coB\"uchi" automaton in the case of "prefix-independent" languages in Section~\ref{subsec-poly:prefix-indep}, and in Section~\ref{subsec-poly:general}, we show how to get rid of the "prefix-independence" assumption.

%In order to simplify the presentation of the minimisation construction and the proof of its optimality, we decompose it in two steps: first we construct a minimal "HD" "generalised coB\"uchi" automaton in the case of "prefix-independent" languages in Section~\ref{subsec-poly:prefix-indep}, and in Section~\ref{subsec-poly:general}, we show how to get rid of the "prefix-independence" assumption.

\subsection{Minimisation of HD coB\"uchi automata}\label{subsec-poly:coBuchi}

\subparagraph{Safe components and safe languages.} 
\AP A path $q\lrpE q'$ in a "coB\"uchi" automaton is ""safe@@path"" if no "coB\"uchi transition" appears on it. 
\AP Let $\intro*\safeA$ be the automaton obtained by removing from $\A$ all "coB\"uchi transitions".
\AP A ""safe component"" of $\A$ is a strongly connected component (i.e., a maximal set of states which are all reachable from each other) of $\safeA$; formally, this is an "automaton structure" $\S=(S,\DD_\S)$ with $S\subseteq Q_\A$ and $\DD_\S\subseteq \DD_\A$. 
\AP We let $\intro*\safeComp{\A}$ be the set of "safe components" of $\A$.

%\begin{remark}\label{rmk:accepting-iff-safeComponent}
%	A "run" in $\A$ is "accepting@@run" if and only if it eventually remains in one "safe component".
%\end{remark}
\AP We define the ""safe language of a state $q$"" as:
\[ \intro*\safeLang{q} = \{ w\in \SS^\oo \mid \text{ there is a "run" } q\lrp{w\phantom{.}} \text { in } \safeA \}. \]

\begin{example}
	The "safe components" of the automaton on the right of Figure~\ref{fig-FP:aut-factor-xx} (page~\pageref{fig-FP:aut-factor-xx}) have as set of states: $S_1 = \{q_0,q_1\}$, $S_2 = \{p_0,p_1\}$ and $S_3 = \{t_0,t_1\}$.
	The "safe language" of $q_0$ is $\safeLang{q_0} = \{w\in \SS^\oo \mid w \text{ does not contain the factor } aa\}$.
\end{example}

% The following statement simply follows from the fact that the "accepting runs" of a "coB\"uchi" automaton are exactly those that eventually stay in a "safe component".

% \begin{lemma}\label{lemma:AK-accepts-iff-safe}
% 	Let $\A$ be a "semantically deterministic" "coB\"uchi" "automaton".
% 	Let $w\in \SS^\oo$ be a word labelling a path in a "safe component" of $\A$ starting from a state $q$. Then, $uw\in L$ for all $u\in \SS^*$ such that $q_\init \lrp{u\phantom{.}} q$.
% 	%labelling a path from the initial state to $q$.
% \end{lemma}

\subparagraph{Nice coB\"uchi automata.}
\AP We say that a "coB\"uchi" automaton $\A$ is in ""normal form"" if all transitions between two different "safe components" are "coB\"uchi transitions". We note that any "coB\"uchi automaton" can be put in "normal form" without modifying its "language@@aut" by setting all transitions between two different "safe components" to be "coB\"uchi@@transition".
\AP We say that $\A$ is ""safe deterministic"" if $\safeA$ is a "deterministic automaton". That is, if the "non-determinism" of $\A$ appears exclusively in "coB\"uchi transitions".
\AP We say that $\A$ is ""nice"" if all its states are "reachable", it is "semantically deterministic", in "normal form", and "safe deterministic".

It is not difficult to see that any "history-deterministic automaton" $\A$ can be assumed to be "semantically deterministic". This means that different choices made by a "resolver" from the same state with different histories must be consistent with the "residual", i.e. must lead to states accepting the same language, which is the language $u^{-1}L(\A)$ after reading a word $u$. Kuperberg and Skrzypczak showed the more involved result that we can moreover assume "safe determinism"~\cite{KS15DeterminisationGFG}. All in all, we have:

\begin{lemma}[\cite{KS15DeterminisationGFG}]\label{lem:nice-coBuchi-automata}
	Every "history-deterministic" "coB\"uchi" "automaton" $\A$ can be turned in polynomial time into an equivalent "nice" "HD" "coB\"uchi" "automaton" $\A_{\mathrm{nice}}$ such that:
	\begin{itemize}
		\item $|\A_{\mathrm{nice}}| \leq |\A|$,
		\item For every "safe component" $\S$ of $\A_{\mathrm{nice}}$, there is some "safe component" $\S'$ of $\A$ with $|\S|\leq |\S'|$.
	\end{itemize}
\end{lemma}

Although the second item of the previous proposition is not explicitly stated in~\cite{KS15DeterminisationGFG}, it simply follows from the fact that all the transformations used to turn $\A$ into a "nice automaton" either add "coB\"uchi" transitions to $\A$ or remove transitions from it. These operations can only subdivide "safe components".

\subparagraph{Minimal HD coB\"uchi automata.} 
We present the necessary conditions  for the minimality of "history-deterministic" "coB\"uchi" automata identified by Abu Radi and Kupferman~\cite{AK22MinimizingGFG}.

\AP We say that a "coB\"uchi" automaton $\A$ is ""safe centralised"" if for all "equivalent states" $q\eqRes p$, if the "safe languages" of $q$ and $p$ are comparable for the inclusion relation ($\safeLang{q}\subseteq \safeLang{p}$ or vice versa), then they are in the same "safe component" of $\A$.
\AP It is ""safe minimal"" if for all states $q\eqRes p$, the equality $\safeLang{q}= \safeLang{p}$ implies $q=p$.

\begin{example}
	The automaton on the right of Figure~\ref{fig-FP:aut-factor-xx} is "safe minimal" and "safe centralised". However, the automaton from Figure~\ref{fig-FP:coBuchi-b-or-c} is not "safe centralised", as $\safeLang{q_1}=\emptyset\subseteq \safeLang{q_2}$, but $q_1$ and $q_2$ appear in different "safe components".
\end{example}

%The following result follows from Lemma~3.5 and Theorem~3.15 in~\cite{AK22MinimizingGFG}.

\begin{lemma}[{\cite[Lemma~3.5]{AK22MinimizingGFG}}]\label{lem:minimal-coBuchi-AK}
	Let $\A_{\min}$ be a "nice", "safe minimal" and "safe centralised" "HD" "coB\"uchi" "automaton". Then, for any "equivalent@@aut" "nice" "HD" "coB\"uchi" automaton $\A$ there is an injection 
	$\intro*\eta \colon \safeComp{\A_{\min}} \to \safeComp{\A}$
	such that for every safe component $\S\in \safeComp{\A_{\min}}$, it holds that $|\S|\leq |\eta(\S)|$.
\end{lemma}

Minimality of such automata directly follows:
\begin{corollary}\label{cor:minimal-coBuchi-AK}
	Let $\A_{\min}$ be a "nice", "safe minimal" and "safe centralised" "HD" "coB\"uchi" "automaton". Then, the number of states of $\A_{\min}$ is minimal among all "HD" "coB\"uchi" automata "recognising" the same language.
\end{corollary}

\begin{theorem}[{\cite[Theorem~3.15]{AK22MinimizingGFG}}]\label{th:computation-minimal-coBuchi-AK}
	\AP Any "coB\"uchi recognisable" language $L$ can be "recognised" by a "nice", "safe minimal" and "safe centralised" "HD" "coB\"uchi" "automaton" $\intro*\autMinAK{L}$.
	Moreover, such an automaton $\autMinAK{L}$ can be computed in polynomial time from any "HD" "coB\"uchi" automaton "recognising" $L$.
\end{theorem}

% \begin{proof}
	% 	Since $\autMinAKL$ is "nice" and $L$ is "prefix-independent", we have that all its states are "equivalent@@state", so in particular, for all $q\in \autMinAKL$, the "language recognised from" $q$ is $L$. 
	% 	Therefore, any word $w\in \SS^\oo$ labelling a path in $\autMinAKL$ not visiting any "coB\"uchi transition" is in $L$, and by "prefix-independence", so does $uw$.
	% \end{proof}

\subsection{Minimal HD generalised coB\"uchi automata: prefix-independent case}\label{subsec-poly:prefix-indep}

In this subsection, we show how to minimise "history-deterministic" "generalised coB\"uchi" automata "recognising" "prefix-independent" languages. The "prefix-independence" hypothesis will be removed in the next subsection. 

Let $L\subseteq \SS^\oo$ be a "prefix-independent" "coB\"uchi recognisable" language, and let $\A$ be a "history-deterministic" "generalised coB\"uchi" automaton "recognising" it.

%\subparagraph*{The construction for prefix-independent languages.}
Combining Corollary~\ref{cor-first:gen-Buchi-to-Buchi} and Theorem~\ref{th:computation-minimal-coBuchi-AK}, we obtain that we can build in polynomial time the minimal "HD" "coB\"uchi" automaton $\autMinAK{L}$ for $L$.
Let $\safeComp{\autMinAKL} = \{\S_1,\dots, \S_k\}$ be an enumeration of the "safe components" of $\autMinAKL$, with $S_i$ and $\DD_i$ the sets of states and transitions of each "safe component", respectively. 
We show how to build an "HD" "generalised coB\"uchi automaton" $\autMinGenCoBPI$ of "size@@aut" $n_{\max} = \max_{1\leq i \leq k} |S_i|$ and using $k$ "output colours".
%In the following, we use the convention $k+1 = 1$.  

Intuitively, $\autMinGenCoBPI$ will be the full automaton (it contains transitions between all pairs of states, for all input letters). 
Since $|S_i|\leq n_{\max}$, we can map each "safe component" $\S_i$ to this full automaton via a "morphism" $\phi_i$, and use (the non-appearance of) colour $i$ to accept "runs" that eventually would have stayed in the "safe component" $\S_i$ in $\autMinAKL$.
That is, the transitions of $\autMinGenCoBPI$ that are ``safe-for-colour $i$'' will be exactly those in $\phi_i(\S_i)$.

\AP Formally, let $\intro*\autMinGenCoBPI = (Q,\SS, q_\init,\DD,\GG,\colAut, \genCoBuchi)$ with:
\begin{itemize}
	\item $Q = \{p_1,p_2,\dots, p_{n_{\max}}\}$,
	\item $q_{\init} = p_1$ (any state can be chosen as "initial@@state"),
	\item $\DD = Q\times \SS \times Q$,
	\item $\GG = \pow{\{1,\dots, k\}}$.
\end{itemize}

%Note that the choice of initial state is not important as we assumed the language to be "prefix-independent".
Finally, we define the "colour labelling" $\colAut\colon \DD \to \GG$. For each $1\leq i\leq k$, let $\phi_i\colon \S_i \to (Q,\DD)$ be any injective "morphism of automaton structures" (such a "morphism" exists since $|\S_i| \leq n_{\max}$ and $(Q,\DD)$ is the full "automaton structure").
We put colour $i$ in a transition $e \in \DD$ if and only if there is no transition $e'\in \DD_i$ such that $\phi_i(e') = e$. That is, $\colAut(e)=\{i \mid \inv{\phi_i}(e) = \emptyset\}$.

%For convenience, we let $\phi_i(q) = p_1$ for all $q\notin \S_i$ (so that $\phi_i\colon Q_{\autMinAKL}\to Q$ is total).

\begin{remark}
	We remark that this "colour labelling" uses some arbitrary choices, namely, the way we map the different "safe components" of $\autMinAKL$ to the full automaton of size $n_{\max}$. In particular, there is no unique minimal "HD" "generalised coB\"uchi" automaton "recognising" $L$.
	By a slight abuse of notation, we denote $\autMinGenCoBPI$ one automaton originated by this procedure.
\end{remark}

\begin{example}
	The automaton on the left of Figure~\ref{fig-FP:aut-factor-xx} (page~\pageref{fig-FP:aut-factor-xx}) (almost) corresponds to this construction. 
	Indeed, it has been obtained by assigning a colour to each "safe component" of $\autMinAKL$ (on the right) and superposing them in a $2$-state automaton.
	To simplify its presentation, we have removed some unnecessary transitions of $\autMinGenCoBPI$, that is why the automaton displayed is not the full-automaton.
\end{example}

%\subparagraph*{Correctness (prefix-independent case).}
\begin{restatable}[Correctness]{proposition}{PropCorrectMinGenCoBPI}\label{prop:correctness-PI}
	Let $L$ be a "prefix-independent" language that is "coB\"uchi recognisable".
	The automaton $\autMinGenCoBPI$ is "history-deterministic" and "recognises" $L$.
\end{restatable}

\begin{proof}[Proof sketch]
	If $w$ admits an "accepting run" $\rr$ in $\autMinGenCoBPI$, then its "run" eventually does not produce some colour $i$ in its "output@@run".
	This means that, eventually, such a run is the $\phi_i$-projection of a "run" in a "safe component" of $\autMinAKL$, so $w\in L$. This is where we use prefix-independence: as soon as there is a witness that a suffix of $w$ is also a suffix of a word in $L$, then the whole word $w$ is in $L$ as well.
	
	A "resolver" for $\autMinGenCoBPI$ can be defined as follows: in a round-robin fashion we follow the different "safe components" of $\autMinAKL$. If a colour $i$ is produced while we are trying to avoid it, we go back to $p_1$ and try to avoid colour $i'=(i+1)\!\mod k$ by following the "safe component" $\S_{i'}$. If a word $w$ belongs to $L$, it eventually admits a "safe path" in $\autMinAKL$, so it will be accepted by this resolver.
\end{proof}

%\subparagraph*{Minimality (prefix-independent case).}

\begin{proposition}[Minimality]\label{prop:minimal-HD-genCoBuchi-prefix-indep}
	Let $\B$ be a "history-deterministic" "generalised coB\"uchi" automaton "recognising" a "prefix-independent" language $L$. 
	Then, $|\autMinGenCoBPI| \leq |\B|$.
	% n_{\max}$, where $n_{\max}$ is the maximal size of a "safe component" of $\autMinAKL$.
\end{proposition}

\begin{proof}
	Let $C = \set{1,\dots,k}$ be the set of "output colours" used by the "acceptance condition" of $\B$ and let $\autGenCoB{C} $ be the "coB\"uchi automaton" "recognising" $\genCoBuchiCol{C}$ presented in Section~\ref{subsec-first:composition}. 
	By \Cref{cor-first:gen-Buchi-to-Buchi}, $\autGenCoB{C}\autComposition \B$ is a "history-deterministic" automaton "recognising" $L$. Moreover, the states of $\autGenCoB{C}\autComposition \B$ are a disjoint union $Q_1\disjUnion Q_2 \disjUnion \dots \disjUnion Q_k$ such that:
	\begin{itemize}
		\item $|Q_i| = |\B|$,
		\item all transitions leaving $Q_i$ are "coB\"uchi transitions" going to $Q_{i+1}$, where $Q_{k+1}=Q_1$.
	\end{itemize}
	Therefore, each "safe component" $\S$ of $\autGenCoB{C}\autComposition \B$ is included in some $Q_i$, so $|\S|\leq |\B|$.
	By Lemma~\ref{lem:nice-coBuchi-automata}, we can turn $\autGenCoB{C}\autComposition \B$ into a "nice" "HD" "coB\"uchi" automaton $\B'$ satisfying that none of its "safe components" is larger than $|\B|$.
	
	By Lemma~\ref{lem:minimal-coBuchi-AK} there is an injection $\eta\colon \safeComp{\autMinAKL} \to \safeComp{\B'}$ such that $|\S|\leq |\eta(\S)|$ for all "safe component" $\S$ of $\autMinAKL$. In particular, if $\S_{\max}$ is a "safe component" of maximal size in  $\autMinAKL$, we obtain:
	$ |\autMinGenCoBPI| = |\S_{\max}| \leq |\eta(\S_{\max})| \leq |\B|.$\qedhere 
\end{proof}

\subsection{Minimal HD generalised coB\"uchi automata: general case}\label{subsec-poly:general}
We now describe the polynomial-time construction for minimising a given "HD" "generalised coB\"uchi" automaton (without the "prefix-independence" assumption). For the optimality proof, we can reduce to the simplest "prefix-independent" case.

%\subparagraph*{The construction (general case).}

We fix an "HD" "generalised coB\"uchi" automaton $\A$ "recognising" a language $L$. 
As before, using Corollary~\ref{cor-first:gen-Buchi-to-Buchi} and the minimisation procedure of Abu Radi and Kupferman, we can obtain the minimal "HD" "coB\"uchi" "automaton" $\autMinAK{L}$ in polynomial time.
We show how to convert it to an "equivalent" "HD" "generalised coB\"uchi" "automaton" $\autMinGenCoB$ of minimal "size@@aut".

Let $R_1,R_2,\cdots,R_m$ be all the distinct "residual languages" of $L$, i.e., languages of the form $\quot{u}{L}$ for some finite word $u \in \Sigma^{*}$. The case $m=1$ corresponds to the "prefix-independent" case treated in Section \ref{subsec-poly:prefix-indep}.
We note that these "residuals" induce a partition of the states of $\autMinAKL$ into $\resStates{R_1},\dots, \resStates{R_m}$, where the states in $\resStates{R_j}$ "recognise" $R_j$.
We assume that $R_1=L$ is the "residual" corresponding to the initial state of $\autMinAKL$.
Let $\S_1, \S_2, \cdots, \S_k$ be the "safe components" of $\autMinAK{L}$, with $S_i$ and $\DD_i$ as sets of states and transitions, respectively. 
For each residual language $R_j$, define $n_j$ as the largest number of states recognising $R_j$ appearing in a "safe component" of $\autMinAK{L}$. That is, 
\vspace{-1.5mm}
\[n_j = \max_{1\leq i \leq k} |S_i\cap \resStates{R_j}|.\]
We shall construct a language-equivalent "HD" "generalised coB\"uchi" automaton $\autMinGenCoB$ with $n_1 + n_2 + \cdots +n_m$ states. Towards this, for each "residual language" $R_j$, let $P_j = \{p_j^1,p_j^2,\cdots,p_j^{n_j}\}$ be a set of $n_j$ elements. 
The automaton $\intro*\autMinGenCoB = (Q,\SS, q_\init, \DD, \GG, \colAut, \genBuchi)$ is given by:

%k nb safe components. i index for them
%m nb residuals. j index

\begin{itemize}
	\item $Q = P_1\disjUnion P_2\disjUnion \cdots \disjUnion P_m$.
	\item $q_\init = p_1^1$ (any state corresponding to the "residual" of the initial state of $\autMinAKL$ would work).
	\item Let $(q,a,q')$ be a transition in $\autMinAKL$, with $q\in \resStates{R_j}$ and $q'\in \resStates{R_{j'}}$. Then, $(p,a,p')\in \DD$ for all $p\in P_j$ and $p'\in P_{j'}$.
	\item $\GG = \pow{\{1,\dots, k\}}$.
\end{itemize}

One way of picturing $\autMinGenCoB$ is by taking the automaton of residuals of $L$ and making $n_j$ copies of the state corresponding to each residual $R_j$ (while keeping all transitions).

We now describe the "colour labelling" $\colAut\colon \DD \to \GG$. Informally, each "safe component" $\S_i$ is mapped into $\autMinGenCoB$ so that the states of $S_i\cap R_j$ are mapped into $P_j$. These "safe components" are then ``superimposed'' upon each other and coloured appropriately, so that a "run" eventually staying in $\S_i$ in $\autMinAKL$ corresponds to a "run" in $\autMinGenCoB$ that eventually avoids colour $i$.

More formally, for $i\in [1,k]$, let $\phi_i\colon \S_i\to \autMinGenCoB$ be an injective "morphism" such that $\phi_i(q) \in P_j$ if $q\in \resStates{R_j}$. 
Such injective "morphism" does indeed exist, by the choice of $n_j$ and the fact that $\autMinGenCoB$ contains all transitions consistent with the residuals.
The transitions of $\DD$ that are $i$-safe are defined to be exactly those that are the image by $\phi_i$ of some transition in $\S_i$.
That is, for $e\in \DD$, the labelling $\colAut(e)$ contains exactly the colours in $\{i \mid \inv{\phi_i}(e) = \emptyset\}$. 

%For convenience, for $q\notin S_i$, $q\in P_j$, we let $\phi_i(q) = p_j^1$.
%As before, some arbitrary choices in this construction for the definition of the $\phi_i$s. We denote $\intro*\autMinGenCoB$ any automaton obtained by this procedure.

%This concludes our description of $\autMinGenCoB$. We claim that it recognises the language $L$, is history-deterministic, and any HD-GCA recognising $L$ must have at least $|\autMinGenCoB|$ many states.

\begin{remark}
	We remark that the automaton $\autMinGenCoB$ obtained in this way uses a polynomial number of "output colours" in the size of the minimal "HD" "coB\"uchi automaton" $\autMinAKL$ (see also \shortlong{long version}{Lemma~\ref{lem-NPHard:bound-nb-colours}}). More precisely, the number $k$ of colours is the number of safe components of $\autMinAKL$. However, this number of colours is not necessarily optimal (see Theorem~\ref{thm-NPHard:min-state-and-col-NPhard}).
\end{remark}

%\subparagraph*{Correctness (general case).}

%We now proof that $\autMinGenCoB$ does indeed "recognise" $L$ and is "history-deterministic".

The correctness of our construction, stated below, is proven similarly to Proposition~\ref{prop:correctness-PI}.

\begin{restatable}[{Correctness}]{proposition}{PropCorrectMinGenCoB}\label{prop:correctness-general}
	Let $L$ be a "coB\"uchi recognisable" language.
	The automaton $\autMinGenCoB$ is "history-deterministic" and "recognises" $L$.
\end{restatable}

In particular, the resolver for $\autMinGenCoB$ is defined as in Proposition~\ref{prop:correctness-PI}: it follows the different "safe components" of $\autMinAKL$ in a round-robin fashion, by trying to avoid colour some colour $i$ while moving in $\phi_i(S_i)$, then if colour $i$ is seen it switches to $i'=(i+1) \mod k$, etc.

%\subparagraph*{Minimality (general case).} 
We explain how to obtain the minimality of $\autMinGenCoB$, stated below. We reduce to the "prefix-independent" case, using a technique from~\cite{BCRV22HalfPosBuchi}.\footnote{An alternative proof scheme is to extend the proof of Proposition~\ref{prop:minimal-HD-genCoBuchi-prefix-indep} to the general case, by taking into account the "residuals" of the language. For this, we need a refinement of Lemma~\ref{lem:minimal-coBuchi-AK}, stating that the injection $\eta$ satisfies that, for every "residual" $R$ and safe component $\S$ of $\autMinAKL$, $|\S\cap R| \leq |\eta(\S) \cap R|$. The proof of Abu Radi and Kupferman~\cite{AK22MinimizingGFG} does indeed lead to this result, but it is not explicitly stated in this form.}
Full proofs can be found in \shortapp{app:minimisation-proofs}.

\begin{restatable}[Minimality]{proposition}{PropMinGeneralCase}\label{prop:minimal-HD-genCoBuchi-general}
	Let $\B$ be a "history-deterministic" "generalised coB\"uchi automaton" "recognising" a language $L$. 
	Then, $|\autMinGenCoB| \leq |\B|$.
\end{restatable}

%\subparagraph*{Localisation at a residual.}
For each "residual" $R=\quot{u}{L}\in \Res{L}$, we define the
\AP ""local alphabet at $R$"", as:
\[ \intro*\locAlph{R} = \{v\in \SS^+ \mid \resClass{uv} = \resClass{u} \text{ and for any proper prefix } v' \text{ of } v, \, \resClass{uv'}\neq \resClass{v} \}. \]

%We note that this definition is independent from the choice of the representative $u$.

That is, if $\A$ is a "semantically deterministic" automaton with states $Q$,
% automaton "recognising" $L$,
 then $\locAlph{R}$ is the set of words that connect states in $\resStates{R}$. 
%We also write $\locAlph{[q]}= \locAlph{R}$ for $q\in \resStates{R}$ a state "recognising@@state" $R$.
%We identify words over $\locAlph{R}$ with words over $\SS$.
%Note that, if it is non-empty, $\locAlph{R}$ is a prefix code, and therefore an infinite word $w\in \SS^\oo$ admits at most one decomposition of the form $w_1w_2\dots$, with $w_i\in \locAlph{R}$. 
Note that in general $\locAlph{R}$ may be infinite, however this is harmless in this context, and we will freely allow ourselves to talk about automata over infinite alphabets.
\AP Also, $\locAlph{R}$ is empty if and only if all the states of $\resStates{R}$ are ""transient"", that is, they do not occur in any cycle of the automaton. 
For simplicity, in the following we assume that no state of $\A$ is "transient"; the general case is treated in detail in \shortapp{app:minimisation-proofs}.
%\AP If this is the case, we say that the "residual" $R$ is ""transient@@res"", on the contrary, we say that it is ""recurrent@@res"".
%\footnotetext{Note that in a finite automata over an infinite alphabet, there are finitely many classes of letters such that two letters from the same class admit exactly the same transitions. Hence one may easily turn such an automaton into an automaton over a finite alphabet.}

%Seeing words in $\locAlph{R}^\omega$ as words in $\SS^\omega$, 
\AP We define the ""localisation of $L$ to"" a "residual"~$R\in \Res{L}$ as the language over the alphabet $\locAlph{R}$ given by:
$
\intro*\locLang{R} = \{ w \in \locAlph{R}^\omega \mid w \in R\}.
$

\begin{remark}
	For every "residual" $R$, $\locLang{R}$ is a "prefix-independent" language. It corresponds to infinite words whose letters are in $\locAlph{R}$ that is, going from $\resStates{R}$ to $\resStates{R}$, and eventually avoid seeing some colour on such paths. The prefix-independence of $\locLang{R}$ follows from the fact that $\locLang{R}$ has only itself as residual, on alphabet $\locAlph{R}$.
\end{remark}

%\begin{remark}
%	For every "recurrent residual" $R$, $\locLang{R}$ is a "prefix-independent" language.
%\end{remark}

Let $\A$ be a "semantically deterministic" "generalised coB\"uchi" automaton with $k$ colours "recognising" $L\subseteq \SS^\oo$.
%, using $W\subseteq \GG^\oo$ as "acceptance condition". 
\AP For each "recurrent residual"~$R$ of $L$ we define $\intro*\localAut{R}$ to be the "generalised coB\"uchi" "automaton" over $\locAlph{R}$ given by:
\begin{itemize}
	\item the set of states is $\resStates{R}$, that is, the set of states of $\A$ "recognising" $R$.
	\item the "initial state" is arbitrary,
	\item the "acceptance condition" is $\genCoBuchiCol{C}$ (for $C$ the "output colours" of $\A$),
	\item there is a transition $q\re{w:X}p$, with $w\in \locAlph{R}$, $X\in \pow{\set{1,\dots,k}}$, if there is a path from $q$ to $p$ labelled $w$ and producing the set of colours $X$ in $\A$.	
\end{itemize}
\begin{restatable}{lemma}{lemLocalisationCorrectness}\label{lem-poly:loc-recognises-loc}
	The automaton $\initialAut{(\localAut{R})}{q}$ "recognises" $\locLang{R}$ for each $q \in \resStates{R}$. Moreover, if $\initialAut{\A}{q}$ is "history-deterministic", so is~$\initialAut{({\localAut{R}})}{q}$.
\end{restatable}

Using  the fact that ("safe@@path") paths between states in $\resStates{R}$ are the same in $\A$ or in $\localAut{R}$, combined with Lemma~\ref{lem:minimal-coBuchi-AK}, we can prove:

%\begin{lemma}\label{lemma-poly:minimal-localisation-AK}
%	$\localAut{u_j}$ is a minimal automaton "recognising" $\locLang{u_j}$. Moreover, a maximal "safe component" of this automaton has size $n_j$.
%\end{lemma}

\begin{lemma}\label{lem-poly:localisation-safeMin-safeCen}
	%For a "residual" $R$ of $L$, the automaton $\localAutA{\autMinAKL}{R}$ is "nice", "safe minimal" and "safe centralised". 
	$\localAutA{\autMinAKL}{R}$ is a minimal "HD" "coB\"uchi" automaton "recognising" $\locLang{R}$. Moreover, a maximal "safe component" of this automaton has size $n_j$.
\end{lemma}

To conclude the proof of Proposition~\ref{prop:minimal-HD-genCoBuchi-general}, we combine \Cref{lem-poly:localisation-safeMin-safeCen} with Proposition~\ref{prop:minimal-HD-genCoBuchi-prefix-indep} to show that  $|\resStatesA{R_j}{\B}|\geq n_j$. This implies:  $|\B| \geq n_1 + \cdots + n_m = |\autMinGenCoB|$. 
%
%This can be easily done by combining \Cref{lem-poly:localisation-safeMin-safeCen} with Proposition~\ref{prop:minimal-HD-genCoBuchi-prefix-indep}.

%
\section{NP-completeness of minimisation of deterministic and HD generalised B\"uchi automata}
\label{sec:NP-hardness}
In this section we contrast the polynomial-time complexity previously obtained for minimising "history-deterministic" "generalised coBüchi" "automata" with the \NP-hardness of the minimisation of "deterministic" "generalised (co)Büchi" and "history-deterministic" "generalised Büchi" "automata".

%We also establish the \NP-completeness of minimising both states and colours for "HD" "coBüchi automata", as well as the three other types of automata mentioned before.

\begin{restatable}{theorem}{ThmMinHDBuchi}
	\label{thm-NPhard:minHDBuchi}
	%The minimisation of number of states of "history-deterministic" "generalised Büchi" "automata" is \NP-complete.
	The following problem is $\NP$-complete:
	Given a "history-deterministic" "generalised B\"uchi" automaton $\A$ and a number $n$, decide whether there is an equivalent "history-deterministic" "generalised B\"uchi automaton" with at most $n$ states.
\end{restatable}

\begin{restatable}{theorem}{ThmMinDetCoBuchi}
	\label{thm-NPhard:mindetcoBuchi}
	The minimisation of the number of states of "deterministic" "generalised B\"uchi" and "deterministic" "generalised coBüchi" "automata" is \NP-complete.
\end{restatable}

We show \NP-hardness of the minimisation problems in the "deterministic" and "history-deterministic" Büchi cases simultaneously. The \NP-hardness for the "deterministic" coBüchi case follows directly.
Our reduction is from a suitable version of the "3-colouring problem". %for graphs. 
%We start by presenting this version of "3-colouring".

We further consider the problem of minimising both colours and states simultaneously for generalised (co)B\"uchi automata.
For the automata classes appearing in the previous theorems ("deterministic" and "history-deterministic" "generalised B\"uchi" automata), it easily follows that this problem is $\NP$-complete.
We focus therefore in the case of "history-deterministic" "generalised coB\"uchi", for which the minimisation of states has proven to be polynomial (Theorem~\ref{th:minimisationHDgenCoBuchi}). We show that, even in this case, minimising both states and colours is $\NP$-complete.

\begin{theorem}\label{thm-NPHard:min-state-and-col-NPhard}
	The following problem is $\NP$-complete:
	Given a "history-deterministic" "generalised coB\"uchi" automaton $\A$, and numbers $n$ and $k$, decide whether there is an equivalent "history-deterministic" "generalised coB\"uchi automaton" with at most $n$ states and $k$ colours.
\end{theorem}

We obtain the lower bound by adapting the proof of \NP-hardness of the minimising of Rabin pairs, given in~\cite{CM24Simplifying}, itself inspired from~\cite{Hugenroth23ZielonkaDAG}.
Details can be found in \shortapp{appendix-NP-hard:min-state-and-colours}.

\subsection{Containment in $\NP$ and bounds on the necessary number of colours}
In this section we address an important subtlety of our minimisation problems: We minimise the number of states, but the number of colours used by a "generalised Büchi" or "generalised coB\"uchi" automaton may be exponential in its number of states. 

In order to show that the problems at hand are in $\NP$, we need to show that the minimal "deterministic" and "history-deterministic" automata require a number of colours that is polynomial in the size of the input (that is, the number of states and colours of the input automaton).
This turns out to be true, although not trivial. 
Full proofs are in \shortapp{appendix-upper-bound}.

\begin{restatable}{lemma}{BoundsColorsDet}
	\label{lem:bound-colors-det}
	Let $\A$ be a "deterministic" "generalised Büchi" "automaton" with $n$ states and $k$ colours. 
	Then there is an "equivalent@@aut" "deterministic" "generalised Büchi" "automaton" with a minimal number of states and using $O(n^2k)$ colours. 
\end{restatable}

\begin{restatable}{lemma}{BoundsColorsHD}
	\label{lem:bound-colors-HD}
	Let $\A$ be a "history-deterministic" "generalised Büchi" "automaton" with $n$ states and $k$ colours. 
	Then there is an "equivalent@@aut" "history-deterministic" "generalised Büchi" "automaton" with a minimal number of states and using $O(n^3k^2)$ colours. 
\end{restatable}

This allows us to obtain an \NP{} algorithm as we only need to guess an automaton with polynomially many states and colours, and check equivalence with the input automaton. The latter test can be done in polynomial time (see for instance~\cite[Thm.~4]{Schewe20MinimisingGFG}).

As an additional result, we show that the previous lemmas do not hold for general "non-deterministic" automata: minimising an automaton may blow up its number of colours.

\begin{restatable}{proposition}{ExpColoursNDGBA}
	There exists a family of "non-deterministic" "generalised Büchi" "automata" $(\A_n)_{n\in \NN}$ such that for all $n$, $\A_n$ uses $n+1$ states and $2$ colours and a minimal automaton equivalent to $\A_n$ requires $2^n$ colours. 
\end{restatable}

\subsection{Hardness of state minimisation}
\label{subsec:np-hardness-state-min}
We provide a reduction from the "3-colouring problem".
We construct from a given graph $G$ a deterministic automaton $\A_G$ such that:
\begin{itemize}
	\item If $G$ is "3-colourable" then there is a 3-state "deterministic" automaton equivalent to $\A_G$, and
	
	\item if $G$ is not "3-colourable" then there is no automaton $\B$ (deterministic or not) with 3 states equivalent to $\A_G$.
\end{itemize}
This establishes the hardness of state-minimisation for "deterministic" and "history-deterministic" automata simultaneously.
The full proof is in \shortapp{appendix-NP-hard-state-min}. We only present here the languages we use and a sketch of the first item. The second item is obtained by a refined case analysis over the cycles of "generalised Büchi" automata with three states.

Given an undirected "graph" $G=(V,E)$, %we say that two vertices $u,v \in V$ are ""neighbours"" if $\set{u,v} \in E$.
we define the ""neighbourhood"" of a vertex $v$ as the set $\intro*\neighbourhood{v} = \set{v' \in V \mid \set{v,v'} \in E}$, and its ""strict neighbourhood"" as $\intro*\neighbourhoodOpen{v} = \neighbourhood{v} \setminus \set{v}$.

We consider the alphabet $\SS = V$. For each $v \in V$, we define the language:
\[L_{v} = (V^* vv)^\oo \cup (V^* (V \setminus \neighbourhood{v}))^\oo \;\;\text{ and we let } \;\; L_G = \bigcap_{v \in V} L_{v}.\]
In words, a sequence of nodes is in $\langGraph$ if for all $v \in V$ it either has infinitely many factors $vv$ or sees a vertex that is not a "neighbour" of $v$ infinitely many times.

The first item is proven by the following more general lemma. We actually show that from a "$k$-colouring" of $G$ we can build a deterministic automaton with $k$ states for $L_{G}$. This also allows us to construct the automaton $\A_G$ by applying this lemma on a trivial $|V|$-colouring.

\begin{restatable}{lemma}{KColourAut}
	\label{lem-NPhard:k-colour-aut}
	For all "graph" $G=(V,E)$ and $k \in \NN$, if $G$ is "$k$-colourable" then there exists a "complete" "deterministic" "generalised Büchi" "automaton" $\B$ with $k$ states which recognises $\langGraph$.
\end{restatable}

\begin{figure}[ht]
	\centering
	\includegraphics[]{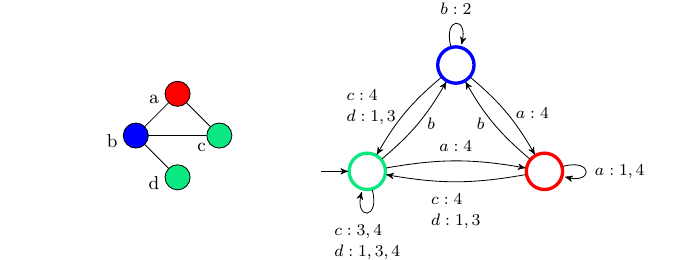}
	\caption{A "graph" with a "3-colouring", and the corresponding automaton as defined in Lemma~\ref{lem-NPhard:k-colour-aut}. The "output colours" $a,b,c,d$ have been replaced by $1,2,3,4$ for readability.}
	\label{fig-NPhard:k-colour-aut}
\end{figure}

\begin{proof}[Proof sketch]
	Suppose $G$ is "$k$-colourable", let $c: V \to \set{1,\ldots, k}$ be a "$k$-colouring" of $G$.	
	We define the "deterministic" "generalised Büchi" "automaton" $\B$ as follows: 
	\begin{itemize}
		\item The set of states is $Q=\set{1,\ldots, k}$, we pick any state as the initial one.
		\item For $q\in Q$ and $v \in \SS=V$, the $v$-transition from $q$ is $q\re{v} c(v)$.
		\item The set of "output colours" is $V$, hence the "output alphabet" is $\GG = \pow{V}$.
		\item If $q\neq c(v)$, the transition  $q\re{v} c(v)$ is coloured with $V\setminus\neighbourhood{v}$. Transitions of the form $c(v)\re{v} c(v)$ are coloured with $V \setminus\neighbourhoodOpen{v} $.
	\end{itemize} 
	
	It is then quite straightforward to show that a word is accepted by $\B$ if and only if for each $v$ it goes infinitely many times through the $v$-loop on $c(v)$ or sees infinitely many times vertices outside of $\neighbourhood{v}$. The structure of the automaton ensures that those words are exactly the ones in $L_v$. In particular, the fact that $c(u) \neq c(v)$ for all "neighbours" $u$ and $v$ implies that we cannot go through the $v$ loop on $c(v)$ without reading a $v$ or a non-neighbour of $v$ just before. 
	Figure~\ref{fig-NPhard:k-colour-aut} shows an example of this construction.
\end{proof}

\section{Conclusion}
\label{sec:conclusion}
We believe that one of the key novel insights of this work is to compare the complexity of the minimisation of "HD" "generalised coB\"uchi" automata (polynomial) with both "HD" "generalised B\"uchi" automata and "deterministic" models ($\NP$-complete).
For "history-deterministic" and "deterministic" "B\"uchi" automata the minimisation problem is still open; our results are an important step in this direction, and seem to indicate that the polynomial-time minimisation algorithm for the "HD" "coB\"uchi" case will not extend to the "B\"uchi" or the "deterministic" case.

\bibliographystyle{plainurl}
\bibliography{references}
\shortlong{}{
\newpage
\appendix
\section{G2 conjecture}
\label{app:G2-conjecture}

To establish that deciding whether a "B\"uchi" automaton is "history-deterministic" can be done in $\PTimeFull$, Bagnol and Kuperberg~\cite{BK18} introduced the "$G_2$ game", played on an automaton $\A$ between two players Eve and Adam.
Polynomial-time decidability is obtained using the facts that (1) the winner of this game can be decided in polynomial time, and, (2) for "B\"uchi" and "coB\"uchi" automata, 
$\A$ is "history-deterministic" if and only if Eve wins "$G_2$" \cite{KS15DeterminisationGFG,BK18,BKLS20SuccRec}.
We discuss now the status of this problem for "generalised B\"uchi" and "generalised coBüchi" automata.

\paragraph*{Definitions and state of the art}

\AP We recall the ""$G_2$ game"" from~\cite{BK18}.
Given an automaton $\A$, the "$G_2$ game" on $\A$, noted $\intro*\Gtwo{\A}$, is played between Eve and Adam in the following way:
The arena is $Q^3$, with starting position $(q_0,q_0,q_0)$.
At each round, from position $(p,q_1,q_2)$:
\begin{enumerate}
	\item Adam chooses a letter $a\in\Sigma$,
	\item Eve chooses a transition $p\re{a}p'$.
	\item Adam chooses transitions $q_1\re{a}q_1'$ and $q_2\re{a}q_2'$.
	\item The game moves to position $(p',q_1',q_2')$
\end{enumerate}
Eve wins the game if either her run $\rho$ on the first component is accepting, or both of Adam's runs $\lambda_1$ and $\lambda_2$ on second and third components are rejecting.

\AP The ""$G_2$ conjecture"", stated in~\cite{BK18} is the following:
\begin{conjecture}
	A parity automaton $\A$ is "history-deterministic" if and only if Eve wins $\Gtwo{\A}$.
\end{conjecture}

So far, the conjecture has been proved for "B\"uchi" automata~\cite[Cor. 21]{BK18} and for "coB\"uchi" automata~\cite[Thm. 28]{BKLS20SuccRec}.

\paragraph*{A general remark: Cascade composition by deterministic automata preserves $G_2$}
We state here a property of the "$G_2$ conjecture" with respect to generic conditions. This idea has been present for some time among researchers in the field,  we explicit it here for clarity.

\begin{lemma}\label{lem:G2cond}
	Let $W_1\subseteq\Sigma^\omega$ and $W_2\subseteq\Gamma^\omega$ be languages on finite alphabets $\Sigma,\Gamma$.
	Assume that the $G_2$ conjecture holds for $W_2$-automata, and that there is a deterministic $W_2$-automaton recognising $W_1$.
	Then the $G_2$ conjecture holds for $W_1$-automata.
\end{lemma}

\begin{proof}
	Let $\D$ be the deterministic $W_2$-automaton recognising $W_1$.
	Let $\A$ be an arbitrary non-deterministic $W_1$-automaton. We want to prove that the $G_2$ conjecture holds for the automaton $\A$.
	We build a non-deterministic $W_2$-automaton $\B=\D\circ\A$.
	Since $\D$ is deterministic, the projection from runs of $\B$ to runs of $\A$ by forgetting the $\D$ component is a bijection, that preserves acceptance. In particular $L(\A)=L(\B)$.
	
	Let us assume that Eve wins $\Gtwo\A$. By ignoring the $\D$ component, Eve can use the same strategy to win $\Gtwo\B$. Since $\B$ is a $W_2$-automaton, by assumption we obtain that $\B$ is HD. The HD strategy in $\B$ can now be used in $\A$, using $\D$ as extra memory. Indeed, Eve can win the letter game of $\A$ by simulating the $\D$ component in her strategy, and play as in the letter game of $\B$.
	We conclude that $\A$ is HD, hence the $G_2$ conjecture holds for $\A$, as for any $W_1$-automaton.
\end{proof}

\paragraph*{Consequences: $G_2$ for generalised (co)B\"uchi automata}

The first consequence that we can notice is that the "$G_2$ conjecture" on parity automata as stated in \cite{BK18} suffices to imply the $G_2$ conjecture for all $\omega$-automata.
This follows from \Cref{lem:G2cond} and the fact that deterministic parity automata are sufficient to recognise any $\omega$-regular language~\cite{Mostowski1984RegularEF}. This feature of the $G_2$ conjecture was already mentioned e.g. in the survey \cite[Sect 6.1]{BL23SurveyHD}.

We finally state another consequence of \Cref{lem:G2cond}.

\begin{theorem}\label{th-FP:G2-conj-genBuchi}
	The "$G_2$ conjecture" holds for "generalised B\"uchi" and "generalised coB\"uchi" automata.
\end{theorem}

\begin{proof}
	Both cases are a simple application of \Cref{lem:G2cond}, as the "$G_2$ conjecture" holds for both "B\"uchi"~\cite[Cor.~21]{BK18} and "coB\"uchi" automata~\cite[Thm.~28]{BKLS20SuccRec}, and every generalised "(co)B\"uchi" condition is recognisable by a deterministic "(co)B\"uchi" automaton (see Section~\ref{subsec-first:composition}).
\end{proof}

Moreover, the $G_2$ game is still tractable for generalised Büchi and coBüchi conditions:
\begin{lemma}\label{lem:G2Ptime}
	The $G_2$ game for generalised Büchi and coB\"uchi automata can be solved in polynomial time.
\end{lemma}

\begin{proof}
%We recall that the simulation game between two automata $\A_1$ and $\A_2$ can be defined similarly as $G_2$, with three differences: 
%\begin{itemize}
%\item Adam plays in $\A_1$ while Eve plays in $\A_2$,
%\item Adam only has one token,
%\item in each round, Adam plays a transition of $\A_1$ before Eve answers it with a transition in $\A_2$.
%\end{itemize}
%The rest is similar, in particular Eve wins if whenever Adam's run in $\A_1$ accepts, Eve's run in $\A_2$ has to accept as well.

	A GR(1) objective is an objective of the form $B_1\Rightarrow B_2$, where $B_1,B_2$ are "generalised B\"uchi" objectives. It is known that GR(1) games can be solved in polynomial time \cite{JuvekarPiterman2006,CDHL16GenBuchiGames}.
	
	The objective of the $G_2$ game of a "generalised coBüchi" automaton is of the form $(C_1\vee C_2)\Rightarrow C_3$, where $C_1,C_2,C_3$ are "generalised coBüchi" objectives.
	Since $C_1\vee C_2$ is still a generalised coBüchi objective, taking the contrapositive yields a GR(1) objective $\neg C_3\Rightarrow (\neg C_1\wedge \neg C_2)$. Thus the $G_2$ game for a coB\"uchi automaton is a GR(1) game, and can be solved in $\PTimeFull$.
	
	For "generalised Büchi" automata, it is a little more complex, because the disjunction of two "generalised Büchi" objectives is not directly a "generalised Büchi" objective.
	However, it can be turned into one with a quadratic blow-up in the number of colours.
	Indeed, this corresponds to performing the following formula expansion:
	$$(\bigwedge_{i\in I} A_i)\vee(\bigwedge_{j\in J}B_j)=\bigwedge_{i\in I}\bigwedge_{j\in J}A_i\vee B_j$$
	Thus, we can turn the objective of a generalised Büchi "$G_2$ game" into a GR(1) objective, where the premise has quadratically many colours, with accepting sets of the form $A_i\cup B_j$.
	This means that as before, solving the $G_2$ game for a generalised B\"uchi automaton can be done in $\PTimeFull$. 
\end{proof}

This gives an alternative proof of \Cref{cor:HDness}.

\section{Correctness and minimality of $\autMinGenCoB$: Proofs for Section~\ref{subsec-poly:general}}\label{app:minimisation-proofs}

We first prove Propositions~\ref{prop:correctness-PI} and~\ref{prop:correctness-general}.

For convenience, we let $\phi_i(q) = p_1$ for all $q\notin \S_i$ (so that $\phi_i\colon Q_{\autMinAKL}\to Q$ is total).

%For convenience, for $q\notin S_i$, $q\in P_j$, we let $\phi_i(q) = p_j^1$.
%As before, some arbitrary choices in this construction for the definition of the $\phi_i$s. We denote $\intro*\autMinGenCoB$ any automaton obtained by this procedure.

\PropCorrectMinGenCoB* 
\begin{proof}
	Let $w\in \Lang{\autMinGenCoBPI}$, and let $\rr\in \DD^\oo$ be an  "accepting run" over $w$. By definition of the "generalised coB\"uchi" acceptance condition, there is a suffix $\rr'$ of $\rr$ and some $i\in [k]$ such that $i$ does not appear in the "output@@run" of $\rr'$. Let $w=uv$ such that $v$ is the label of $\rr'$, and $q$ be the state reached after $u$ in $\rr$. By definition of the "labelling" $\colAut$ of $\autMinGenCoBPI$, this means that $\rr'$ is the $\phi_i$-projection of a path in $\S_i$, starting in some state $p\in Q^{R_j}$ for some $j$. Since $v$ is accepted from $p$ in $\autMinAKL$, we have that $v\in R_j$. Notice that by construction of $\autMinAKL$, the state $q=\phi_i(p)$ is in $P_j$. Since $\autMinAKL$ is semantically deterministic, any state $p'$ reached by $u$ from the initial state of $\autMinAKL$ is in $Q^{R_j}$ as well. We can conclude that $w=uv$ is accepted by $\autMinAKL$, so $w\in L$. This shows that $\Lang{\autMinGenCoBPI}\subseteq L$.
	%We will now show the existence of a resolver for ${\autMinGenCoBPI}$ accepting any word from $L$, thereby completing the proof that ${\autMinGenCoBPI}$ is an HD automaton recognising $L$.

	We now define a "resolver" $\ss\colon \SS^* \times \SS \to \DD$ for ${\autMinGenCoBPI}$ accepting any word from $L$, thereby completing the proof that ${\autMinGenCoBPI}$ is an HD automaton recognising $L$. %in such a way that, for every $w\in L$, the "run induced by this resolver" over $w$ is "accepting@@run". 
	The "resolver" will try to follow the different "safe components" of $\autMinAKL$ in a round-robin manner, by using $k$ memory states.
	Let $\ss_{0}$ be a "resolver" for $\autMinAKL$. 
	We first remark that, by "safe determinism" of $\autMinAKL$, for all state $q$ in $\autMinGenCoBPI$ and for all $a\in \SS$, there is at most one transition $(q,a,q')$ in $\phi_i(\S_i)$, for each $i$.
	The "resolver" $\ss$ will use $k$ memory states. Assume that we have read so far $u\in \SS^*$, reaching a state $q$, and letter $a\in \SS$ is given. If the resolver is in the $i^\mathrm{th}$ memory state, $\ss$ will indicate to take, if it exists, the only $a$-transition from $q$ available in $\phi_i(\S_i)$. In this case, the memory state of $\ss$ is not updated. If, on the other hand, no such transition exists, then $\ss$ will choose the transition $(q,a,\phi_{i+1}(\resolvPath{\ss_0}(ua)))$, and update its memory state to $i+1$ (or $1$, if $i=k$). That is, we look at the state reached in $\autMinAKL$ following the "resolver" $\ss_0$, and jump to its $i+1$-projection in $\autMinGenCoBPI$ and start simulating the run as if we where in the $(i+1)^\mathrm{th}$ "safe component" $S_{i+1}$.
	We show that $\ss$ builds an "accepting run" whenever the input word is in $L$. Let $w\in L$, and let $\rr$ be the "accepting run" over $w$ built by the "resolver" $\ss_0$ in $\autMinAKL$. 
	Eventually, $\rr$ will stay in a "safe component" $\S_j$. Let $N$ be a large enough index such that the suffix of $\rr$ after position $N$ does not leave $\S_j$.
	Suppose by contradiction that the "run on" $w$ built by $\ss$ in $\autMinGenCoBPI$ was "rejecting@@run". Then, it sees all "output colours" in $\{1,\dots,k\}$ infinitely often, therefore, it leaves each component $\phi_i(\S_i)$ infinitely many times.
	However, whenever this run enters $\phi_j(\S_j)$ after position $N$, it will coincide with the $\phi_j$-projection of $\rr'$, so it will not leave this component, a contradiction.
\end{proof}

We now provide all details for the proof of minimality of $\autMinGenCoB$.

\PropMinGeneralCase*

%
%To obtain the minimality of $\autMinGenCoB$ we will reduce to the "prefix-independent" case.\footnote{An alternative proof scheme is to extend the proof of Proposition~\ref{prop:minimal-HD-genCoBuchi-prefix-indep} to the general case, by taking into account the "residuals" of the language. For this, we need a refinement of Lemma~\ref{lem:minimal-coBuchi-AK}, stating that the injection $\eta$ satisfies that, for every "residual" $R$ and safe component $\S$ of $\autMinAKL$, $|\S\cap R| \leq |\eta(\S) \cap R|$. The proof of Abu Radi and Kupferman~\cite{AK22MinimizingGFG} does indeed lead to this result, but it is not explicitly stated in this form.}
%The full proofs can be found in Appendix~\ref{}

\subparagraph*{Localisation at a residual.}
We recall that the "local alphabet at" a "residual" $R=\quot{u}{L}\in \Res{L}$ is:
\[ \locAlph{R} = \{v\in \SS^+ \mid \resClass{uv} = \resClass{u} \text{ and for any proper prefix } v' \text{ of } v, \, \resClass{uv'}\neq \resClass{v} \}. \]
We note that this definition is independent from the choice of the representative $u$.
%If $\A$ is a "semantically deterministic" automaton "recognising" $L$, we also write $\locAlph{[q]}= \locAlph{R}$ for $q\in \resStates{R}$ a state "recognising@@state" $R$.

We identify words over $\locAlph{R}$ with words over $\SS$.
Note that $\locAlph{R}$ is a prefix code, meaning that a word in $\locAlph{R}$ can never be a strict prefix of another word in $\locAlph{R}$. Therefore an infinite word $w\in \SS^\oo$ admits at most one decomposition of the form $w_1w_2\dots$, with $w_i\in \locAlph{R}$. 
\AP As noted before, $\locAlph{R}$ is empty if and only if all the states of $\resStates{R}$ are "transient", that is, they do not occur in any cycle of the automaton. 
\AP If this is the case, we say that the "residual" $R$ is ""transient@@res"", on the contrary, we say that it is ""recurrent@@res"".

For $R$ a "recurrent residual" we let:
\[
\locLang{R} = \{ w \in \locAlph{R}^\omega \mid w \in R\},
\]
which is a "prefix-independent" language.

Also, for a "recurrent residual"~$R$ of $L$, the automaton $\localAut{R}$ is the "generalised coB\"uchi" "automaton" over $\locAlph{R}$ obtained by restricting $\A$ to the states "recognising" $R$.
%We also denote this automaton $\localAut{[q]}$, for $q$ any state of $\A$ "recognising@@state" $R$.

%Let $\A$ be a "semantically deterministic" "generalised coB\"uchi" automaton with $k$ colours "recognising" $L\subseteq \SS^\oo$.
%%, using $W\subseteq \GG^\oo$ as "acceptance condition". 
%\AP For each "recurrent" "residual"~$R$ of $L$ we define $\localAut{R}$ to be the "generalised coB\"uchi" "automaton" over $\locAlph{R}$ given by:
%\begin{itemize}
%	\item the set of states is $\resStates{R}$, that is, the set of states of $\A$ "recognising" $R$.
%	\item the "initial state" is arbitrary,
%	\item the "acceptance condition" is $\genCoBuchiCol{C}$ (for $C$ the "output colours" of $\A$),
%	\item there is a transition $q\re{w:X}p$, with $w\in \locAlph{R}$, $X\in \pow{\set{1,\dots,k}}$, if there is a path from $q$ to $p$ labelled $w$ and producing the set of "colours" $X$ in $\A$.	
%\end{itemize}
%
%
The following lemma directly follows from the definition.

\lemLocalisationCorrectness*

\subparagraph*{Safe components and safe languages of the localisation.} 

\begin{lemma}\label{lem-poly:safe components-localisation}
	Let $\A$ be a "semantically deterministic" "coB\"uchi" automaton. Two states $q,p\in \resStates{R}$ are in a same "safe component" in $\A$ if and only if they are in a same "safe component" in $\localAut{R}$. 
	Also, for all $q\in \resStates{R}$, the "safe language" of $q$ in $\localAut{R}$ is $\safeLang{q}\cap \locAlph{R}^\oo$. 
\end{lemma}

\begin{proof}
	Follows from the fact that ("safe@@path") paths between states in $\resStates{R}$ are the same in $\A$ or in $\localAut{R}$ (note that here $\localAut{R}$ is a "coBüchi automaton").
\end{proof}

Lemma~\ref{lem-poly:localisation-safeMin-safeCen} follows from the following Lemma, combined with Lemma~\ref{lem:minimal-coBuchi-AK}. 
%in the case where $\A$ may contain "transient" states.

\begin{lemma}\label{lem-poly-app:localisation-safeMin-safeCen}
	If $R$ is a "recurrent residual" of $L$, then $\localAutA{\autMinAKL}{R}$ is "nice", "safe minimal" and "safe centralised".
	If $R$ is "transient@@res", then $\autMinAKL$ contains a single state "recognising@@state" $R$.
\end{lemma}
\begin{proof}
	Recall that by Theorem~\ref{th:computation-minimal-coBuchi-AK}, $\autMinAKL$ is "nice", "safe minimal" and "safe centralised". 
	
	First, assume that $R$ is "transient", and let $q$ be a state of $\autMinAKL$ "recognising@@state" $R$. Since $\autMinAKL$ is in "normal form" and all transitions outgoing from $q$ change of strongly connected component, the "safe language" of $q$ is the empty set. If there is $p\eqRes q$, $p\neq q$, the state $p$ would appear in a different "safe component" of $\autMinAKL$, contradicting the fact that it is "safe centralised".
	
	Assume that $R$ is "recurrent@@res". It is a direct check that $\localAutA{\autMinAKL}{R}$ is "nice".
	"Safe minimality" and "safe centrality" follow from Lemma~\ref{lem-poly:safe components-localisation}.
\end{proof}

We can finally combine previous lemmas together with Proposition~\ref{prop:minimal-HD-genCoBuchi-prefix-indep} to obtain minimality of $\autMinGenCoB$ in the general case.

\begin{proof}[Proof of Proposition~\ref{prop:minimal-HD-genCoBuchi-general}]		
	Let $\B$ be an "HD" "generalised coB\"uchi" automaton "recognising" $L$. 
	We use the notations introduced at the beginning of this subsection: $\Res{L}=\{R_1,\dots, R_m\}$, and $n_j$ is the maximal number of states corresponding to $R_j$ appearing in a same "safe component" of $\autMinAKL$.

	We claim that $|\resStatesA{R_j}{\B}|\geq n_j$. This will conclude the proof, as it implies:
	\[ |\B| \geq n_1 + \cdots + n_m = |\autMinGenCoB|. \]
	First, if $R_j$ is a "transient residual", by Lemma~\ref{lem-poly:localisation-safeMin-safeCen}, $n_j=1$, so the inequality holds.
	
	Assume that $R_j$ is "recurrent@@res". By Lemmas~\ref{lem-poly:loc-recognises-loc} and~\ref{lem-poly:localisation-safeMin-safeCen}, 
	$\localAutA{\autMinAKL}{R_j}$ is a minimal "HD" "coB\"uchi" automaton "recognising" $\locLang{R_j}$ (which is a "prefix-independent" language). 
	Moreover, by Lemma~\ref{lem-poly:safe components-localisation},  $\localAutA{\autMinAKL}{R_j}$ contains a "safe component" of size $n_j$. Therefore, by Proposition~\ref{prop:minimal-HD-genCoBuchi-prefix-indep},  an "HD" "generalised coB\"uchi" automaton "recognising" $\locLang{R_j}$ has at least $n_j$ states.
	We have, by Lemma~\ref{lem-poly:loc-recognises-loc}, that $\localAutA{\B}{R_j}$ is an "HD" "generalised coB\"uchi" automaton "recognising" $\locLang{R_j}$, so we conclude that 
	$|\resStatesA{R_j}{\B}| = |\localAutA{\B}{R_j}|\geq n_j$, as we wanted.
\end{proof}
\section{\NP-completeness of minimisation for (history-)deterministic generalised Büchi automata: Proofs for Section~\ref{sec:NP-hardness}}

\subsection{Upper bound: Proof of Proposition~\ref{prop-NPhard:det-and-HD-in-NP}}
\label{appendix-upper-bound}

%\DetHDInNP*
\begin{proposition}%{DetHDInNP}
	\label{prop-NPhard:det-and-HD-in-NP}
	The minimisation of "deterministic" "generalised Büchi" and "generalised coB\"uchi" and "history-deterministic" "generalised Büchi" automata are in $\NP$.
\end{proposition}

We establish the \NP-upper bounds for state minimisation of "deterministic" and "history-deterministic" "generalised Büchi automata".
The result for "deterministic" "generalised coBüchi automata" follows by duality (Remark~\ref{rmk-FP:duality-Buchi-coBuchi}).

We start by establishing our key technical lemma, from which the proposition will follow. It states that the number of colours needed by a "history-deterministic" "Büchi" "automaton" is polynomial in its size and the size of a minimal equivalent "deterministic" "Büchi" "automaton".

\AP We say that we can ""recolour"" a "generalised (co)Büchi automaton" $\A$ with $k$ colours if we can replace its "colour labelling" with one using $k$ colours without changing the language.%The propositions below allow us to "recolour" a minimal automaton with a polynomial number of colours in the input.

\AP A "history-deterministic" "automaton" is called ""resolver-trim"" if there is a resolver $\ss$ for it such that for every state $q$ there is an "accepting run" "induced by" $\ss$ that goes through $q$.
Note that the notions of "resolver-trim" matches the notion of trim over "deterministic" "automata".

\begin{lemma}
	\label{lem-NPHard:bound-nb-colours}
	Let $\A$ be a "resolver-trim" "history-deterministic" "generalised Büchi" automaton with $n_\A$ states. If there exists an "equivalent@@aut" "deterministic" "Büchi" automaton $\B$ with $n_\B$ states, then one can "recolour" $\A$ to obtain an "equivalent@@aut" "resolver-trim" "history-deterministic" "generalised Büchi" automaton with $n_\A$ states and using at most $n_\A n_\B$ colours. 
\end{lemma}
\begin{proof}
	Let $\A = (Q_\A, \SS, q_{\init}^\A, \DD_\A, \GG_\A, \colAut_\A, \genBuchi_{\GG})$ be a "resolver-trim" "history-deterministic" "generalised Büchi" automaton with $n_\A = \size{Q_\A}$ states and using $k = \size{\GG}$ colours. Let $\ss$ be a resolver for $\A$ such that every state is visited by an "accepting run" following $\ss$.
	Let $\B= (Q_\B, \SS, q_{\init}^\B, \DD_\B, \set{1}, \colAut_\B, \genBuchi_{\set{1}})$ be a "deterministic" "Büchi" "automaton" with $n_\B = \size{Q_\B}$ states such that $\Lang{\A} = \L(\B)$.
	
	We prove the lemma by showing that either we can remove a colour from $\A$ or $k \leq n_\A n_\B$. For all $i \in \GG_\A$ let $\A_{-i}$ be the automaton obtained by removing the colour $i$ from $\A$, i.e., composing $\colAut_\A$ with a projection over $\GG_\A \setminus \set{i}$ and replacing the acceptance condition by $\genBuchi_{\GG\setminus\set{i}}$.
	Note that every "accepting" "run" in $\A$ is still accepting in $\A_{-i}$.
	Hence we must have $\Lang{\A} \subseteq \Lang{\A_{-i}}$.
	
	\begin{itemize}
		\item First suppose there exists $i$ such that $\Lang{\A} = \Lang{\A_{-i}}$. As every "accepting" "run" in $\A$ is still accepting in $\A_{-i}$, $\ss$ is still a "resolver" for $\A_{-i}$. 
		As a result, $\A_{-i}$ is a "resolver-trim" "history-deterministic" "generalised Büchi" automaton with the same states and transitions as $\A$ and using $k-1$ colours.
		
		\item Now suppose that for all $i \in \GG_\A$ we have $\Lang{\A} \subsetneq \Lang{\A_{-i}}$. 
		As $\Lang{\A_{-i}}\setminus \Lang{\A}$ is $\oo$-regular and non-empty, it contains an ultimately periodic word, which itself has an ultimately periodic "accepting run" in $\A_{-i}$.
		As a consequence, we can find a state $q_i^\A$ and words $u_i$, $v_i$, such that $u_iv_i^\oo$ is not accepted by $\A$ and there exists a path $q_{\init}^\A \xrightarrow{u_i} q_i^\A$ and a cycle $q_i^\A \xrightarrow{v_i} q_i^\A$ whose output contains all colours of $\GG_\A$ except $i$.

		We show that $k \leq n_\A n_\B$. Suppose by contradiction that $k > n_\A n_\B$. By the pigeonhole principle, there exists $q^\A \in Q_\A$ such that at least $n_\B +1$ distinct colours $i \in \GG_{\A}$ satisfy $q_i^\A = q^\A$. Let $I = \set{i\in \GG_{\A} \mid q_i^\A = q^\A}$.
		
		By definition of $\ss$, there is a word $u\in \SS^*$ such that the "path induced by" $\ss$ when reading $u$ from $q_{\init}^\A$ ends in $q^\A$. For all $i \in I$, $u v_i^\oo$ cannot be in $\Lang{\A}$, as otherwise $\A$ would contain an "accepting run" over $v_i^\oo$ from $q_\A$, meaning that $u_i v_i^\oo$ would also be in $\Lang{\A}$, a contradiction. Hence $u v_i^\oo$ is not in $\Lang{\A}$ for any $i \in I$.
		
		Let $q^\B$ be the state reached in $\B$ by reading $u$ from $q_{\init}^\B$.
		For each $i\in I$, we can find a state $q_i^\B$ in $\B$ and $\alpha_i, \beta_i > 0$ such that there is a path $q^\B \xrightarrow{v_i^{\alpha_i}} q_i^{\B}$ and a cycle $q_i^\B \xrightarrow{v_i^{\beta_i}} q_i^{\B}$. Furthermore, as $u v_i^\oo$ is not accepted by $\A$, it is not accepted by $\B$ and thus there are no Büchi transitions along that cycle.
		
		By the pigeonhole principle, there  exist $i\neq j$ such that $q_i^\B = q_j^\B$. As a result, the word $u v_i^{\alpha_i} (v_i^{\beta_i} v_j^{\beta_j})^\oo$ is not accepted by $\B$: its run in $\B$ goes to $q^\B$ and then goes indefinitely through the cycles reading $v_i^{\beta_i}$ and $v_j^{\beta_j}$, which, as we saw, contain no Büchi transition.
		
		However, $u v_i^{\alpha_i} (v_i^{\beta_i} v_j^{\beta_j})^\oo$ has an accepting run in $\A$: we can read $u$ to arrive in $q_\A$, and then read every factor $v_i$ (resp. $v_j$) by going from $q_\A$ to itself and seeing every colour but $i$ (resp. $j$). As $i\neq j$, this run sees every colour infinitely many times. This is a contradiction, as we assumed $\Lang{\A} = \L(\B)$.\qedhere
	\end{itemize} 
\end{proof}

This lemma allows us to "recolour" the minimal automata with a polynomial number of colours in the input.

\BoundsColorsDet*

\begin{proof}
	By Corollary~\ref{cor-first:gen-Buchi-to-Buchi} there exists a "deterministic" "Büchi" "automaton" $\A'$ equivalent to $\A$ and with $nk$ states.	
	Let $\B$ be a minimal "deterministic" "generalised Büchi" "automaton" equivalent to $\A$. By minimality, $\B$ has at most $n$ states, hence by Lemma~\ref{lem-NPHard:bound-nb-colours} we can "recolour" it using at most $n^2k$ colours. As "recolouring" preserves determinism, we obtain the lemma. 
\end{proof}

\BoundsColorsHD*

\begin{proof}
	By Corollary~\ref{cor-first:gen-Buchi-to-Buchi} there exists a "history-deterministic" "Büchi" "automaton" $\A'$ equivalent to $\A$ and with $nk$ states.
	Using the quadratic determinisation construction for "history-deterministic" "Büchi" "automata"~\cite[Theorem~8]{KS15DeterminisationGFG}, we can construct a "deterministic" "Büchi" "automaton" $\A''$ equivalent to $\A$ and with at most $O(n^2k^2)$ states.	
	Let $\B$ be a minimal "history-deterministic" "generalised Büchi" "automaton" equivalent to $\A$. By minimality, $\B$ has at most $n$ states and is "resolver-trim", hence by Lemma~\ref{lem-NPHard:bound-nb-colours} we can "recolour" it using at most $O(n^3k^2)$ colours, yielding the lemma. 
\end{proof}

We can now conclude the proof of Proposition~\ref{prop-NPhard:det-and-HD-in-NP}.

\begin{proof}[Proof of Proposition~\ref{prop-NPhard:det-and-HD-in-NP}]
	We consider the cases of "deterministic" and "history-deterministic" automata separately:
	
	\textbf{For deterministic automata:}
	Given a bound $n$ and a deterministic "generalised Büchi" automaton $\A$ with $n_\A$ states and $k_\A$ colours, one can guess a deterministic "generalised Büchi" automaton $\B$ with at most $n$ states and $O(n_\A^2 k_\A)$ colours.
	It suffices then to check that $\A$ and $\B$ are "equivalent@@aut". This can be done in polynomial time by turning both automata into "equivalent@@aut" "deterministic" "Büchi" ones (Corollary~\ref{cor-first:gen-Buchi-to-Buchi}) and then checking "equivalence@@aut" between the resulting "deterministic" "Büchi" "automata", which can be done in polynomial time~\cite{ClarkeDK93Unified}.
	
	\textbf{For history-deterministic automata:}
	Given a bound $n$ and a "HD" "generalised Büchi" automaton $\A$ with $n_\A$ states and $k_\A$ colours, one can guess an "HD" "generalised Büchi" automaton $\B$ with at most $n$ states and $O(n_\A^3 k_\A^2)$ colours.
	It suffices then to check that $\A$ and $\B$ are "equivalent@@aut". This can be done in polynomial time by turning both automata into "equivalent@@aut" "HD" "Büchi" ones (Corollary~\ref{cor-first:gen-Buchi-to-Buchi}) and then checking "equivalence@@aut" between the resulting "HD" "Büchi" "automata", which can be done in polynomial time via a simulation game, see for instance~\cite[Thm.~4]{Schewe20MinimisingGFG}\footnote{Schewe's proof concerns state-based automata, but the exact same proof works for transition-based ones. One can also use his work as a black box by translating both automata into equivalent state-based ones (with an at most quadratic blow-up) and then using Schewe's method to check equivalence.}. 
\end{proof}

We then prove that the number of colours may increase exponentially when minimising a non-deterministic "generalised Büchi" automaton. A symmetric proof shows that it is also the case for non-deterministic "generalised coBüchi" automata.

\ExpColoursNDGBA*

\begin{proof}
	We define $\A_n$ as follows:
	Its set of states is $Q = \set{q_{\init}, q_1, \ldots, q_n}$, with $q_{\init}$ the initial state, its input alphabet is $\Sigma= \set{1, \ldots, 2n}$, and its output alphabet is $\GG = \set{\aa, \bb}$.
	For all $i, j \in \set{1,\ldots,n}$ there are transitions $q_{\init} \xrightarrow{i : \varepsilon} q_j$, $q_{i} \xrightarrow{2i-1 : \alpha} q_i$ and $q_{i} \xrightarrow{2i : \beta} q_i$.
	There are no other transitions.
	This automaton recognises the language 
	\[\set{w \in \Sigma^\oo \mid \exists i \in \set{1,\ldots, n}, \set{2i-1, 2i}\subseteq \minf(w)}.\]
	It can also be described as the set of words $w$ such that $\minf(w)$ satisfies $\phi = (1 \land 2) \lor \ldots \lor (2n-1 \land 2n)$, where each letter is interpreted as $\top$ if it is in the set and $\bot$ otherwise.
	This DNF formula can be put in CNF, but the resulting formula has exponentially more clauses:
	
	\[\bigwedge_{i_1 \in \set{1,2}}\cdots \bigwedge_{i_n \in \set{2n-1,2n}} (i_1 \lor \ldots \lor i_n).\]
	
	From this observation we can build an "equivalent@@aut" ("deterministic") "generalised Büchi" "automaton" with a single state: it suffices to assign a colour to each of those clauses, and for each letter $i \in \Sigma$ put a $i$-loop labelled by the colours corresponding to clauses containing $i$.

	Moreover we cannot use less than $2^n$ colours for an equivalent automaton with one state: 
	Suppose there exists such an automaton, let $\GG$ be its set of output colours. For each $\gamma \in \GG$ let $\SS_\gg$ be the set of letters $a$ such that $\gg$ appears in a loop reading $a$.
	The language of that automaton is then the set of words $w$ such that $\minf(w)$ satisfies $\bigwedge_{\gg \in \GG} \bigvee_{i \in \SS_\gg} i$.
	If $\size{\GG} < 2^n$, then we have a CNF formula equivalent to $\phi$ with less than $2^n$ clauses. It is folklore that such a formula does not exist, but we reprove it here for completeness.
	
	Let $\psi$ be a CNF formula over $\SS$ equivalent to $\phi$.
	Let $V \subseteq \pow{\SS}$ be the set of sets of colours which contain exactly one of $2i-1$ or $2i$ for each $i\in \set{1,\ldots, n}$. 
	Each clause in $\psi$ must contain either $2i-1$ or $2i$ for each $i\in \set{1,\ldots, n}$, as otherwise $\set{2i-1, 2i}$ would not satisfy $\psi$.
	As a consequence, each clause is satisfied by all sets of $V$ except at most one.
	As $\psi$ is equivalent to $\phi$, which is satisfied by no set of $V$, $\psi$ must have at least $2^\GG$ clauses: each one allowing to rule out one of the sets of $V$.
	
	In conclusion, every one-state automaton "equivalent@@aut" to $\A_n$ must use at least $2^n$ colours. 
\end{proof}
\subsection{NP-completeness of state minimisation: Proof of Theorems~\ref{thm-NPhard:minHDBuchi} and~\ref{thm-NPhard:mindetcoBuchi}}
\label{appendix-NP-hard-state-min}

We start by presenting a version of the "3-colouring" that will be used to show $\NP$-hardness of the minimisation of (history-)deterministic "B\"uchi" automata.

\paragraph*{Colourings of triangle-full 4-clique-free graph}

\AP A ""finite undirected graph"" is a pair $G = (V,E)$ with $V$ a finite set of \emph{vertices} and $E \subseteq \binom{V}{2}$ a set of pairs of vertices, called \emph{edges}. In all that follows we will simply use the term \reintro{graph} for a "finite undirected graph". We say that two vertices $u,v \in V$ are ""neighbours"" if $\set{u,v} \in E$.

\AP Recall that we defined the \reintro{neighbourhood} of a vertex $v$ as the set $\neighbourhood{v} = \set{v' \in V \mid \set{v,v'} \in E}$, and its \reintro{strict neighbourhood} as the set $\neighbourhoodOpen{v} = \neighbourhood{v} \setminus \set{v}$.
Let $k \in \NN$, a ""$k$-clique"" is a set of $k$ vertices $\set{v_1, \ldots, v_k}$ such that $\set{v_i, v_j} \in E$ for all $i\neq j \in \set{1,\ldots, k}$. 

\AP We say that a "graph" is ""triangle-full"" if all vertices are in a "3-clique". It is ""4-clique-free"" if it does not have any "4-clique".

\AP A ""$k$-colouring"" of a "graph" $G$ is a function $c: V \to \set{1,\ldots, k}$ such that for all $\set{v,v'} \in E$, $c(v) \neq c(v')$.
A "graph" is ""$k$-colourable"" if there exists a "$k$-colouring" of it. 
The ""colouring problem"" asks, given a "graph" and an integer $k$, if that "graph" is "$k$-colourable". 
The "3-colouring problem" asks if a given "graph" is "3-colourable". 
Those problems are well known to be \NP-complete~\cite{Stockmeyer73Graph3colour}. They are in fact already \NP-complete for "triangle-full" "4-clique-free" "graphs":

\begin{lemma}
	\label{lem-NPhard:triangle-full}
	The following problem is \NP-hard: 
	Given a "triangle-full" "4-clique-free" "graph", is it "3-colourable"?
\end{lemma}

\begin{proof}
	We reduce from the general "3-colouring problem". Let $G = (V,E)$ be a "graph". We define a "triangle-full" "4-clique-free" "graph" $\Tilde{G}$ such that $G$ is "3-colourable" if and only if $\Tilde{G}$ is.
	
	If $G$ contains a "4-clique" then it is not "3-colourable". We can check whether this is the case in polynomial time, and if so, we we can simply set $\Tilde{G}$ as an arbitrary "triangle-full" "4-clique-free" non-"3-colourable" "graph", for instance the one in Figure~\ref{fig-NPhard:red-triangle-full}.
	
	Otherwise we define the "graph" $\Tilde{G} = (V \times \set{0,1,2}, E_0\cup E_{012})$, such that $E_0 = \set{\set{(v,0),(v',0)} \mid \set{v,v'} \in E}$ and $E_{012} = \set{\set{(v,i),(v,j)} \mid v \in V, i\neq j \in \set{0,1,2}}$.

	\begin{figure}[H]
		\centering
		\includegraphics[]{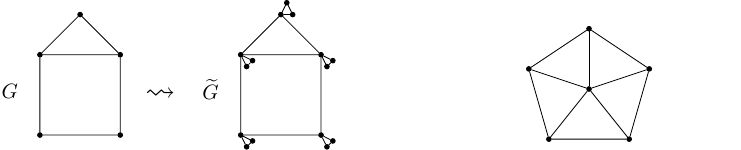}
		\caption{On the left, an example of the reduction of Lemma~\ref{lem-NPhard:triangle-full}. On the right, a "triangle-full" "4-clique-free" "graph" that is not "3-colourable".}
		\label{fig-NPhard:red-triangle-full}
	\end{figure}
	
	Clearly $\Tilde{G}$ is "triangle-full", as $(v,0), (v,1), (v,2)$ are connected to each other for all $v$. Furthermore, it does not contain a "4-clique" as $G$ does not and all additional vertices have only two "neighbours".
	
	Further, if $\Tilde{G}$ is "3-colourable", then so is $G$: it suffices to colour each vertex $v$ in $G$ with the colour of $(v,0)$ in $\Tilde{G}$.
	Conversely, if $G$ is "3-colourable", then we can colour every vertex $(v,0)$ in $\Tilde{G}$ with the colour of $v$ in $G$, and then colour $(v,1)$ and $(v,2)$ with the two remaining colours.
	
	As a result, $G$ is "3-colourable" if and only if $\Tilde{G}$ is.
\end{proof}

\paragraph*{The reduction: Automata and graph colourings}

We now associate with each "triangle-full", "4-clique-free" "graph" $G =(V,E)$ a language $\langGraph$ such that $\langGraph$ is recognised by a 3-state "generalised Büchi" "automaton" if and only if $G$ is "3-colourable".

For each $v \in V$, we define 
\[\intro*\langVertex{v} = (V^* vv)^\oo \cup (V^* (V \setminus \neighbourhood{v}))^\oo \;\;\text{ and we let } \;\; \intro*\langGraph = \bigcap_{v \in V} \langVertex{v}.\]
In other words, a word over $V$ is in $\langGraph$ if for all $v \in V$ it either has infinitely many factors $vv$ or sees a vertex that is not a "neighbour" of $v$ infinitely many times.

We first prove that, given a "graph" $G$, we can build in polynomial time a deterministic "generalised B\"uchi" automaton "recognising" $\langGraph$.

\begin{lemma}
	\label{lem:PTIME-graph-Lv}
	Given a "graph" $G=(V,E)$, we can build in polynomial time a "deterministic" "generalised B\"uchi" automaton "recognising" $\langGraph$.
\end{lemma}
\begin{proof}
	We define a  "generalised B\"uchi" automaton over the alphabet $\SS = V$.
	Automaton $\A$ is given by:
	\begin{itemize}
		\item The set of states is $Q=V$, we pick any state as the initial one.
		\item For $q\in Q$ and $v\in \SS=V$, the $v$-transition from $q$ is $q\re{v} v$.
		\item The set of "output colours" is $V$, hence the "output alphabet" is $\GG = \pow{V}$.
		\item If $q\neq v$, the transition  $q\re{v} v$ is coloured with $V\setminus\neighbourhood{v}$. Transitions of the form $v\re{v} v$ are coloured with $V \setminus\neighbourhoodOpen{v} $.
	\end{itemize}
	That is, the automaton uses the structure of the "graph"; when reading a vertex $v$ we jump to the state corresponding to $v$, and produce as output the colours corresponding to vertices that are not on its "neighbourhood". Self-loops do moreover produce colour $v$. 
	
	Therefore, each time that either a vertex not in $\neighbourhood{v}$ or the factor $vv$ is read, colour $v$ is produced, and these are the only cases in which colour $v$ is produced. Therefore, $\Lang{\A}=\langGraph$. 
\end{proof}

We then extend the previous lemma: from a "$k$-colouring" of the "graph" we can construct an automaton with $k$ states recognising $\langGraph$.

\KColourAut*

\begin{proof}
	Suppose $G$ is "$k$-colourable", let $c: V \to \set{1,\ldots, k}$ be a "$k$-colouring" of $G$.	
	We define the "deterministic" "generalised Büchi" "automaton" $\B$ as follows: 
	\begin{itemize}
		\item The set of states is $Q=\set{1,\ldots, k}$.
		\item For $q\in Q$ and $v\in \SS=V$, the $v$-transition from $q$ is $q\re{v} c(v)$.
		\item The set of "output colours" is $V$, hence the "output alphabet" is $\GG = \pow{V}$.
		\item If $q\neq v$, the transition  $q\re{v} v$ is coloured with $V\setminus\neighbourhood{v}$. Transitions of the form $v\re{v} v$ are coloured with $V \setminus\neighbourhoodOpen{v} $.
	\end{itemize} 
	
	We argue that $\L(\B) = \langGraph$. Let $w = v_0 v_1 \cdots \in \langGraph$, as $\B$ is "deterministic" and "complete" $w$ has a unique "run" in $\B$. 
	
	Let $v \in V$, $w$ is in $\langVertex{v}$. We show that the "run" of $\B$ over $w$ must see colour $v$ infinitely often.
	We distinguish two cases:
	\begin{itemize}
		\item If $vv$ appears infinitely often as a factor of $w$, then by definition of $\B$, after reading the first $v$ the "run" reaches state $c(v)$. As a result, the transition reading the second $v$ must be $c(v) \xrightarrow{v} c(v)$, which is labelled by $V \setminus \neighbourhoodOpen{v}$. In particular the colour $v$ is seen infinitely often. 
		
		\item Otherwise, there is a vertex $u \notin \neighbourhood{v}$ that is read infinitely often. As $u \notin \neighbourhood{v}$, $v \notin \neighbourhood{u}$, thus $v$ appears in the label of all transitions reading $u$. As a result, the colour $v$ is seen infinitely often.
	\end{itemize}
	
	We have shown that the "run" of $\B$ over $w$ sees each label of $\GG_G$ infinitely often, hence the "run" is "accepting". As a consequence, $\langGraph \subseteq \L(\B)$.
	
	For the other inclusion, let $w = v_0 v_1 \cdots \in \L(\B)$, and let $\rr$ be an "accepting run" of $w$ in $\B$. 
	For each $v \in V$, the label $v$ is seen infinitely often in that "run".
	We show that $w$ must then be in $\langVertex{v}$.	
	The transitions labelled by $v$ are the ones reading vertices that are not "neighbours"  of $v$ and the loop on $c(v)$ reading $v$. 
	We distinguish two cases:
	\begin{itemize}
		\item If $w$ contains infinitely many occurrences of vertices of $V\setminus\neighbourhood{v}$ then it is in $\langVertex{v}$. 
		
		\item Otherwise, $w$ ultimately only contains vertices of $\neighbourhood{v}$. As colour $v$ is seen infinitely often in $\rho$, the loop $c(v) \xrightarrow{v} c(v)$ must be taken infinitely often.
		Furthermore, the transitions going to $c(v)$ are all labelled by vertices mapped to $c(v)$ by $c$. As $c$ is a "$k$-colouring" of $G$, apart from $v$ none of those vertices can be in $\neighbourhood{v}$.
		Therefore, eventually every execution of the loop $c(v) \xrightarrow{v} c(v)$ must be preceded by a transition reading $v$, which implies that the factor $vv$ appears infinitely many times in $w$, and thus that $w$ is in $\langVertex{v}$.
	\end{itemize}
	
	We showed that $w$ is in $\langVertex{v}$ for all $v$, hence $\L(\B) \subseteq \langGraph$.
	We have proven that $\B$ recognises $\langGraph$, which concludes our proof.
\end{proof}

We must now show the other implication, that an "automaton" with 3 states recognising $\langGraph$ induces a "3-colouring" of $G$. To do so, we will rely on the notions of "$v$-cycle" and "$\neighbourhoodOpen{v}$-cycle".

%%%AC: I remove the environment definition because is the only one in the paper
Let $G = (V,E)$ be a "graph", and $\A$ an "automaton" over the alphabet $V$.
\AP A ""$v$-cycle"" (resp. ""$\neighbourhoodOpen{v}$-cycle"") $C$ in $\A$ is a non-empty set of states such that for all $q, q' \in C$, there is a non-empty path from $q$ to $q'$ with only transitions reading $v$ (resp. letters in $\neighbourhoodOpen{v}$).

\begin{lemma}
	\label{lem-NPhard:aut-to-colour}
	For all "triangle-full" "4-clique-free" "graph" $G =(V,E)$, if there exists a ("non-deterministic") "complete" "generalised Büchi" "automaton" $\B$ with three states recognising $\langGraph$ then $G$ is "3-colourable".
\end{lemma}

\begin{proof}
	We will proceed in two main steps: First we will show that for all $v \in V$ there is exactly one state $q_v$ that is part of a "$v$-cycle" (Claim~\ref{claim-NPhard:single-q-SCC}). Then we will show that two "neighbours" $u, v$ cannot share that state, i.e., $q_u \neq q_v$ (Claim~\ref{claim-NPhard:cu-neq-cv}).
	As $\B$ has three states, mapping each vertex $v$ to $q_v$ yields a "3-colouring" of $G$.
	
	At many points we will use the fact that as $\B$ is a "generalised Büchi" "automaton" in the following way. If we have an "accepting run" $\rr$ and a "run" $\rr'$ such that every transition occurring infinitely often in $\rr$ also occurs infinitely often in $\rr'$, then $\rr'$ is accepting: as $\rr$ sees all colours infinitely many times, so does $\rr'$.
	
	Let us start with the following observation.
	
	\begin{claim}
		\label{claim-NPhard:q-out-of-n-SCC}
		For all $v \in V$, there exists $q \in Q$ such that $q$ is not in a "$\neighbourhoodOpen{v}$-cycle".  
	\end{claim}
	
	\begin{claimproof}
		Suppose by contradiction that each $q \in Q$ is in a "$\neighbourhoodOpen{v}$-cycle". Then there exists a word $w_q \in \neighbourhoodOpen{v}^+$ and a path from $q$ to $q$ reading $w_q$. 	
		Let $\set{u_1, \ldots, u_k} = \neighbourhoodOpen{v}$.
		The word $(u_1^2\cdots u_k^2 v^2)^\oo$ is in $\langGraph$, as for all $u \in V$ either this word contains infinitely many $uu$ or $u$ is not a "neighbour" of $v$, which appears infinitely many times. 
		Consider an accepting "run" $\rr$ over this word. We transform $\rr$ by inserting after each $v$ a path from $q$ to itself reading $w_q$, where $q$ is the state reached after reading that $v$. 
		
		We obtain a run reading a word in which all vertices read are in the "neighbourhood" of $v$, but which contains no $vv$ factor, hence is not in $\langGraph$. However, all edges seen infinitely often in the first run are also seen infinitely often in the second one, thus the latter is accepting, a contradiction.
	\end{claimproof}
	
	This first claim allows us to proceed with the first step of the proof: We argue in the following claim that for all vertex $v$, at most one state is in a "$v$-cycle".

	\begin{claim}
		\label{claim-NPhard:single-q-SCC}
		For all $v \in V$ there is one and only one state $q_v \in Q_\B$ that is part of a "$v$-cycle". 
	\end{claim}
	
	\begin{claimproof}
		As we assumed that $\B$ is "complete", there must exist a "$v$-cycle" for all $v$ (by reading $v$'s from any state we eventually encounter a cycle). The difficulty is then to show that two distinct states cannot both be in "$v$-cycles".
		
		Let $\set{q_0, q_1, q_2} = Q_\B$. 
		Suppose by contradiction that there exists $v \in V$ such that two distinct states are in "$v$-cycles". 
		Without loss of generality we assume that it is the case for $q_0$ and $q_1$.
		As $G$ is "triangle-full", $v$ has two "neighbours" $t$ and $u$ such that $\set{t,u}\in E$. Furthermore, as $G$ is "4-clique-free", $t$, $u$ and $v$ have no common "neighbour", hence the word $(t^2u^2v^2)^\oo$ is in $\langGraph$.
		Let $\rr$ be an accepting "run" reading that word in $\B$.
		As $q_0$ and $q_1$ are in "$v$-cycles", for all $i \in \set{0,1}$ there exists $k_i > 0$ and a path $\pi_i$ from $q_i$ to itself reading $v^{k_i}$.
		We construct a run $\rr'$ by inserting in $\rr$, at every occurrence of $q_0$ and $q_1$, the paths $\pi_0$ and $\pi_1$, respectively.
		As $\rr$ is accepting and all transitions taken infinitely often in $\rr$ also appear infinitely often in $\rr'$, $\rr'$ is accepting. 
		Hence it must read a word of $\L(\B) = \langGraph$. 
		As it only reads vertices $t$, $u$ and $v$, in particular it must read $t^2$ and $u^2$ infinitely often. By construction of $\rr'$, the state reached between two consecutive $t$'s or $u$'s must be $q_2$, as otherwise a path reading $v$s would have been inserted.
		
		However, there cannot be a loop on $q_2$ reading $t$, or two transitions reading $t$ to and from $q_i$ for some $i \in \set{0,1}$, as otherwise $q_2$ would be in a "$t$-cycle", and thus all three states would be in a "$\neighbourhoodOpen{u}$-cycle", contradicting Claim~\ref{claim-NPhard:q-out-of-n-SCC}.
		The only possibility is that for some $i\in \set{0,1}$, there is a transition $q_i \xrightarrow{t} q_2$ and another one $q_2 \xrightarrow{t} q_{1-i}$, and no other transition reading $t$ to or from $q_2$. 
		By the same argument, for some $j\in \set{0,1}$, there is a transition $q_j \xrightarrow{u} q_2$ and another one $q_2 \xrightarrow{t} q_{1-j}$, and no other transition reading $u$ to or from $q_2$.
		
		\begin{figure}[ht]
			\centering
			\includegraphics[]{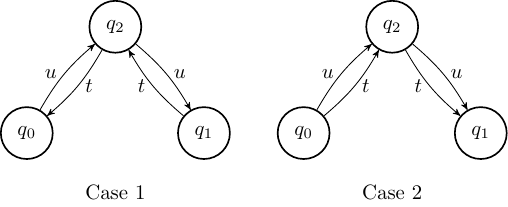}
			\caption{An illustration of the case distinction from the proof of Claim~\ref{claim-NPhard:single-q-SCC}}
			\label{fig-NPhard:single-q-scc}
		\end{figure}
		
		\emph{Case 1:} $i\neq j$. Then the transitions reading $t$ and $u$ induce a "$\neighbourhoodOpen{v}$-cycle" covering all states, contradicting Claim~\ref{claim-NPhard:q-out-of-n-SCC}.
		
		\emph{Case 2:} $i=j$. Then we obtain a run $\rr''$ by applying the following transformation to $\rr'$. We showed that two consecutive $t$ read in $\rr'$ could only be read through the transitions $q_i \xrightarrow{t} q_2 \xrightarrow{t} q_{1-i}$. we replace each occurrence of those transitions with $q_i \xrightarrow{t} q_2 \xrightarrow{u} q_{1-i}$, which  we can do as $i=j$.
		Similarly, we replace each occurrence of $q_i \xrightarrow{u} q_2 \xrightarrow{u} q_{1-i}$ with $q_i \xrightarrow{u} q_2 \xrightarrow{t} q_{1-i}$.
		
		As both $tt$ and $uu$ are read infinitely often in $\rr'$, and the runs $\rr'$ and $\rr''$ see the same set of transitions infinitely often, hence $\rr''$ is accepting. However, it is easy to see that $\rr''$ never reads two consecutive $u$ but only reads vertices from $\set{t,u,v} \subseteq \neighbourhood{u}$, thus $\rr''$ is an accepting "run" of $\B$ reading a word that is not in $\langGraph$, a contradiction.
		
		We reached a contradiction in both cases, proving the claim. 
	\end{claimproof}
	
	By Claim~\ref{claim-NPhard:single-q-SCC}, we can define a mapping $c : V \to Q_\B$ such that for all $v\in V$, $\set{c(v)}$ is the only "$v$-cycle". It remains to show that $c$ is a "3-colouring".
	
	\begin{claim}
		\label{claim-NPhard:cu-neq-cv}
		For all $(u,v) \in E$, $c(u) \neq c(v)$.
	\end{claim}
	
	\begin{claimproof}
		Suppose by contradiction that $c(u) = c(v)$.
		Let $\set{u_1, \ldots, u_k} = \neighbourhoodOpen{v} \setminus \set{u}$.
		The word $(u^2v^2u_1^2\cdots u_k^2)^\oo$ is in $\langGraph$, and thus there is an accepting "run" $\rr$ reading it.
		As $\set{c(u)}$ is the only "$u$-cycle" in $\B$, after reading $u^2$ the reached state can only be $c(u)$ (as $\B$ has three states). Moreover, by reading $v$ from $c(u) = c(v)$ we stay in $c(v)$, as $\set{c(v)}$ is the only "$v$-cycle". 
		
		Hence after each factor $u^2v$, the run $\rr$ reaches $c(v)$. We construct $\rr'$ by inserting additional transitions $c(v) \xrightarrow{u} c(v)$ at each occurrence of $c(v)$ in $\rr$. As $\rr$ is "accepting@@run" and the set of transitions occurring infinitely often in $\rr'$ contains the ones of $\rr$, $\rr'$ is "accepting@@run".	
		However, $\rr'$ reads only letters from $\neighbourhood{v}$, and never reads two consecutive $v$'s, thus it does not read a word of $\langGraph$, a contradiction.
	\end{claimproof}
	
	By Claim~\ref{claim-NPhard:cu-neq-cv}, the map $c$ is a "3-colouring" of $G$, hence $G$ is "3-colourable", proving the lemma.
\end{proof}

We can now conclude the proofs of Theorems~\ref{thm-NPhard:minHDBuchi} and~\ref{thm-NPhard:mindetcoBuchi}.

\ThmMinHDBuchi*

\begin{proof}
	The upper bound follows from Proposition~\ref{prop-NPhard:det-and-HD-in-NP}.
	For the lower bound, we have shown that the "3-colouring problem" is \NP-hard for "triangle-full" "4-clique-free" "graphs".
	Furthermore, given a "triangle-full" "4-clique-free" "graph" $G$, we can construct in polynomial time an automaton $\A_G$ recognising a language $L_G$ (Lemma~\ref{lem:PTIME-graph-Lv}) such that:
	\begin{itemize}
		\item If $G$ is "3-colourable" then there is a "deterministic" (hence also "history-deterministic") "generalised Büchi" automaton $\B$ with three states "recognising" $L_G$ (Lemma~\ref{lem-NPhard:k-colour-aut}).
		
		\item If there is an "generalised Büchi" automaton $\B$ with three states "recognising" $L_G$ (in particular, if there is a "history-deterministic" one), then $G$ is "3-colourable" (Lemma~\ref{lem-NPhard:aut-to-colour}). 
	\end{itemize}

	Hence the "3-colouring problem" over "triangle-full" "4-clique-free" "graphs" reduces to the minimisation problem for "HD" "generalised Büchi" automata, which is therefore \NP-hard. 
\end{proof}

\ThmMinDetCoBuchi*

\begin{proof}
	We only need to prove the "generalised Büchi" case, as \NP-completeness for "generalised coB\"uchi" automata follows by duality (Remark~\ref{rmk-FP:duality-Buchi-coBuchi}).
	We proceed similarly as in the previous proof.
	
	The upper bound is again given by Proposition~\ref{prop-NPhard:det-and-HD-in-NP}.
	For the lower bound, we have shown that the "3-colouring problem" is \NP-hard for "triangle-full" "4-clique-free" "graphs".
	Furthermore, given a "triangle-full" "4-clique-free" "graph" $G$, we can construct in polynomial time an automaton $\A_G$ recognising a language $L_G$ (Lemma~\ref{lem:PTIME-graph-Lv}) such that:
	\begin{itemize}
		\item If $G$ is "3-colourable" then there is a "deterministic" "generalised Büchi" automaton $\B$ with three states "recognising" $L_G$ (Lemma~\ref{lem-NPhard:k-colour-aut}).
		
		\item If there is a "generalised Büchi" automaton $\B$ with three states "recognising" $L_G$ (in particular, if there is a "deterministic" one), then $G$ is "3-colourable" (Lemma~\ref{lem-NPhard:aut-to-colour}). 
	\end{itemize}
	
	Hence the "3-colouring problem" over "triangle-full" "4-clique-free" "graphs" reduces to the minimisation problem for "deterministic" "generalised Büchi" automata, which is therefore \NP-hard. 
\end{proof}

%	We reduce from the "3-colouring problem" for "triangle-full" "4-clique-free" "graphs", which is \NP-complete by Lemma~\ref{lem-NPhard:triangle-full}. 
%	
%	Let $G = (V, E)$ be a "triangle-full" "4-clique-free" "graph". As $G$ is trivially "$|V|$-colourable", by Lemma~\ref{lem-NPhard:k-colour-aut}, we can construct a "deterministic" "generalised Büchi" "automaton" $\A_G$ with $|V|$ states recognising $\langGraph$. Clearly $\A_G$ can be constructed in polynomial time.
%	
%	By Lemmas~\ref{lem-NPhard:k-colour-aut}, if $G$ is "3-colourable" then there is a "deterministic" (hence also "history-deterministic") "generalised Büchi" "automaton" with three states equivalent to $\A_G$.
%	By Lemma~\ref{lem-NPhard:aut-to-colour}, if there is a "generalised Büchi" "automaton" with three states equivalent to $\A_G$ (thus in particular if there is a "deterministic" or "history-deterministic" one), then $G$ is "3-colourable".
%	
%	This concludes our hardness proof.

%\begin{corollary}
%	\label{cor-NPhard:mindetcoBuchi}
%	The minimisation of "deterministic" "generalised coBüchi" "automata" is \NP-hard.
%\end{corollary}
%
%\begin{proof}
%	Every "deterministic" "generalised Büchi" (resp. coBüchi) "automaton" recognising a language $L$ can be seen as a "deterministic" "generalised coBüchi" (resp. Büchi) "automaton" recognising $\SS^\oo \setminus L$.
%	Hence the minimisations of "deterministic" "generalised Büchi" and "generalised coBüchi" "automata" are actually the same problem.
%\end{proof}

\subsection{$\NP$-completeness for minimising states and colours}\label{appendix-NP-hard:min-state-and-colours}

We consider the problems of minimising both colours and states simultaneously and show that it is $\NP$-complete for (history-)deterministic generalised (co)B\"uchi automata.
For deterministic "generalised B\"uchi" automata, "deterministic" "generalised coB\"uchi" automata, and for "history-deterministic" "generalised B\"uchi" automata, we are able to lift our proof of $\NP$-hardness for state-minimisation to $\NP$-hardness for minimising of states and colours simultaneously, without much difficulty.

\begin{theorem}\label{thm-minsc:trivial}
    The following problems are $\NP$-complete:
    \begin{enumerate}
        \item Given a "deterministic" "generalised B\"uchi" automaton $\A$ and numbers $n$ and $k$, is there an equivalent "deterministic" "generalised B\"uchi" automata with at most $n$ states and $k$ colours?
        \item Given a "deterministic" "generalised coB\"uchi" automaton $\A$ and numbers $n$ and $k$, is there an equivalent "deterministic" "generalised coB\"uchi" automata with at most $n$ states and $k$ colours?
        \item Given a "history-deterministic" "generalised B\"uchi" automaton $\A$ and numbers $n$ and $k$, is there an equivalent "history-deterministic" "generalised B\"uchi" automata with at most $n$ states and $k$ colours?
    \end{enumerate}
\end{theorem}
\begin{proof}
	Containment in $\NP$ for each of these problems is not quite immediate as if $k$ is encoded in binary, a witness automaton could have exponentially many colours in the size of the input.
	However, by Lemmas~\ref{lem:bound-colors-det} and~\ref{lem:bound-colors-HD} we know that this cannot happen: if there is an equivalent automaton $\B$ with $n$ states and $k$ colours in the same class of automata, then there is one with polynomially many colours in the number of states and colours of $\A$. One can thus guess $\B$ and check for simulation between $\A$ and $\B$ both ways to verify language equivalence~\cite{Schewe20MinimisingGFG} and whenever applicable, "history-determinism"~\cite[Theorem 4.1]{HP06}.

    The $\NP$-hardness for minimising states and colours for "deterministic" and "history-deterministic" "generalised B\"uchi" automata  follows from the proofs of Lemmas~\ref{lem-NPhard:k-colour-aut} and~\ref{lem-NPhard:aut-to-colour}, from where we can infer that there is a "deterministic" "generalised B\"uchi" automaton $\B$ with $k$ states and $k$ colours recognising $L_G$ if and only if $G$ is "$k$-colourable".

    By duality in the "deterministic" case, this also yields $\NP$-hardness of state and colour minimisation for "deterministic" "generalised coB\"uchi" automata.
\end{proof}

For proving the $\NP$-hardness of state and colour minimisation for HD generalised coB\"uchi automata (\Cref{thm-NPHard:min-state-and-col-NPhard}), we reduce from the "colouring problem" for graphs. 

In all that follows we will only consider connected graphs, and assume that every vertex has degree at least $2$. It is easy to see that the "colouring problem" remains $\NP$-complete with those assumptions.

We select an arbitrary vertex $v_{\init} \in V$.
\AP A ""pseudo-path"" in $G$ is a (finite or infinite) sequence $v_0 e_0 v_1 e_1 \cdots\in (V\cup E)^\infty$ such that $v_i, v_{i+1} \in e_i$ for all $i$. Note that we allow $v_i$ and $v_{i+1}$ to be equal, hence the term "pseudo-path"; that is, we allow a "pseudo-path" to step on an edge without going through it, and come back to the previous vertex. 
\AP A "pseudo-path" is ""initial@@pseudo"" if $v_0 = v_{\init}$. We say that such a "pseudo-path" ""stabilises around"" $v$ if it is infinite and there exists $i$ such that for all $j>i$, $v_j = v$, i.e., the "pseudo-path" eventually stays on the same vertex and just steps on the adjacent edges.
\AP We write $\intro*\Stab{v}$ for the set of "initial@@pseudo" "pseudo-paths" "stabilising around" $v$, and we define the language $\intro*\Lstab{G} = \bigcup_{v \in V} \Stab{v}$ over the alphabet $V \cup E$.
\AP For $v\in V$, we write $\intro*\adj{v}$ for the set of edges $\set{e \in E \mid v\in e}$.

We define the "automaton" $\A_{G}$ as follows:
\begin{itemize}\setlength\itemsep{0.5mm}
	\item $Q =  V \cup E \cup \{q_\init\}$, where $q_\init$ is a fresh element, which is the initial state,
	\item $\SS = V \cup E$,
	\item $\Delta = \{(q_\init,v_\init,v_\init)\} \cup \set{(v, e,  e) \in V \times E^2 \mid v \in e} \cup \set{(e, v,  v) \in E \times V^2 \mid v \in e}$,
	\item the colour of a transition reading $v \in V$ is $V \setminus \set{v}$, the colour of a transition reading $e = \set{v_1, v_2}$ is $V \setminus \set{v_1, v_2}$	
\end{itemize}

This automaton is "deterministic" and "recognises" the language $\bigcup_{v \in V} Stab(v)$. Note that it is not "complete": it only reads "initial@@pseudo" "pseudo-paths" of $G$.
It can easily be made complete by adding a sink state.

\begin{lemma}
	\label{lem:NP-hard-colours-cannot-reduce-states}
	The automaton $\A_G$ has an equivalent "history-deterministic" "generalised coBüchi automaton" $\B$ with at most $|Q|$ states and $k$ colours if and only if $\A_G$ can be "recoloured" with $k$ colours.
\end{lemma}

\begin{proof}
	The right-to-left direction is immediate. We now prove the left-to-right direction.
	
	We assumed that every vertex in $G$ has degree at least $2$. As there is at most one edge between two vertices. As a consequence, $adj(v) \neq adj(v')$ for all $v \neq v' \in V$.
	This implies that all states of $\A_G$ have distinct residuals.
	
	As $\B$ is history-deterministic, for each residual of the language, it must have a state recognising that residual. Hence $\B$ has exactly $|Q|$ states, one for each residual.
	It is therefore deterministic, and its states and transitions are isomorphic to $\A$.
	
	We can thus simply copy the colouring function of $\B$ on the transitions of $\A$ to obtain an equivalent "generalised coBüchi automaton".
\end{proof}

\begin{lemma}
	A  connected "graph" $G$ admits a "$k$-colouring@@graph" if and only if $\A$ can be "recoloured" with $k$ colours.
\end{lemma}
\begin{proof}
	
	For the left-to-right direction, let $c \colon V \to \{1,...,k\}$ be a "$k$-colouring@@graph" of $G$. We define the colouring $col' : \Delta \to \pow{\set{1,...,k}}$ with $col'(e \xrightarrow{v} v) = \set{1,...k} \setminus \set{c(v)}$ and $col'(v_1 \xrightarrow{e} e) = \set{1,...k} \setminus \set{c(v_1), c(v_2)}$ for all $e = \set{v_1, v_2} \in E$.
	Let us prove that $\L(\A) = \L(\A')$.
	Let $w \in \L(\A)$, $w$ is an "initial@@pseudo" "pseudo-path" and there must exist $v$ such that $w$ "stabilises around" $v$. Let $i = c(v)$, ultimately $w$ only visits $v \cup adj(v)$ and thus it never produces colour $i$, so $w \in \L(\A')$.
	Let $w \in \L(\A')$. Again, $w$ is an "initial@@pseudo" "pseudo-path" and there must exist $i$ such that $w$ never sees $i$ after some point, meaning that it ultimately only visits $\bigcup_{v\in c^{-1}(i)} \set{v} \cup adj(v)$. 
	Moreover, as $c$ is a "$k$-colouring@@graph" of $G$, $c^{-1}(i)$ is an independent set (that is, no two vertices are connected by an edge), hence $w$ cannot visit infinitely often two distinct vertices from this set without visiting infinitely often an intermediate vertex of a different colour.
	As a consequence, $w$ must "stabilise around" some $v \in c^{-1}(i)$, thus $w \in \L(\A)$. We have shown that $\L(\A) = \L(\A')$.
	
	For the other direction of the reduction, suppose we have a colouring $col' : \Delta \to \pow{\set{1,...,k}}$ such that  $\A'$ obtained by replacing $col$ with $col'$ in $\A$ is equivalent to $\A$.
	Then we define a colouring $c : V \to \set{1,...,k}$ as follows. For all $v\in V$ we choose $c(v) \notin \bigcup_{e \in E, v \in e} col'(e \xrightarrow{v} v) \cup col'(v \xrightarrow{e} e)$. It is well-defined as seeing all those transitions infinitely many times yields a run reading a word in $\Stab{v}$, hence this "run" must be "accepting" and thus must avoid a colour.
	%We then show that we cannot have $\env(u) \subseteq \env(v)$ for any two neighbours $u$ and $v$.
	
	Two neighbours $u,v$ cannot be coloured with the same colour $i$ by $c$, as otherwise one could read the word $u_v (e v e v')^\oo $ with $u_v$ a pseudo-path ending in $v$ and $e = \set{v,v'}$. This word is not in the language of $\A_G$, but reading it would yield a "run" eventually avoiding colour $i$, hence an "accepting" "run", a contradiction.
	
	We thus obtain that $c$ is a "$k$-colouring@@graph" of $G$.
\end{proof}

By Lemma~\ref{lem:NP-hard-colours-cannot-reduce-states}, $G$ has a "$k$-colouring" if and only if $\A$ can be "recoloured" with $k$ colours without changing its language, if and only if there is an automaton with $ \leq |Q|$ states and $\leq k$ colours equivalent to $\A$.

}

\end{document}